\documentclass[12pt]{article}

\usepackage[utf8]{inputenc}
\usepackage[T1]{fontenc}
\usepackage{amsmath, amssymb, amsfonts}
\usepackage{graphicx}
\usepackage{float}
\usepackage{setspace}
\usepackage{enumitem}
\usepackage{authblk}
\usepackage{parskip}
\usepackage{xcolor}
\usepackage{siunitx}
\usepackage{hyperref}
\usepackage{geometry}
\usepackage{natbib}
\usepackage{booktabs}
\usepackage{caption}
\usepackage{subcaption}
\usepackage{tikz}
\definecolor{linkcolor}{RGB}{0, 90, 160}
\hypersetup{
    colorlinks=true,
    linkcolor=linkcolor,
    urlcolor=linkcolor,
    citecolor=linkcolor
}

\renewcommand{\P}{\mathbb{P}}

\title{Kicking for Goal or Touch? An Expected Points Framework for Penalty Decisions in Rugby Union}
\author{Kenny Watts\thanks{Corresponding author. \href{mailto:landowatts@gmail.com}{landowatts@gmail.com}.}}
\author{Jonathan Pipping-Gam\'on\thanks{ORCID \href{https://orcid.org/0009-0000-2540-2469}{0009-0000-2540-2469}.}}
\affil{\small The Wharton School, University of Pennsylvania, Philadelphia, PA, USA}
\date{\small Published in the \emph{Journal of Quantitative Analysis in Sports} 22(4), 407--421 (2026).\\
\href{https://doi.org/10.1515/jqas-2025-0183}{doi:10.1515/jqas-2025-0183}}

\begin{document}

\maketitle

\begin{abstract}
Following a penalty in rugby union, teams typically choose between attempting a kick at goal or kicking to touch to pursue a try. We develop an \emph{Expected Points} (EP) framework that quantifies the value of each option as a function of field location and game context. Using phase-level data and observed penalty kicks, we construct two context-aware surfaces: (i) the expected points of a possession beginning with a lineout and (ii) the expected points of attempting a kick at goal. Comparing these surfaces yields decision maps that identify where kicking for goal or kicking to touch maximizes expected points, and how the boundary shifts with game context and expected meters gained to touch. To our knowledge, this is the first comprehensive EP-based assessment of penalty strategy in rugby union; it also provides a foundation for future refinement with richer event data and extension to win-probability analysis.
\end{abstract}

\section{Introduction}

Rugby union presents a recurring strategic decision after a penalty. Teams can either attempt an immediate three-point kick at goal or pursue a potential seven-point try via a scrum, a tap-and-go (both from the mark), or a kick to touch followed by a lineout. Except when very close to the opposition try line, teams generally prefer the lineout option to a tap-and-go or scrum because the kick to touch can safely gain territory before the restart. As a result, once a side is within realistic penalty-kick range (roughly 65\,m from the opposition try line), the practical decision is usually between kicking at goal or kicking to touch.

While the choice is often straightforward near the opposition corner, for a given scoreline, or late in a match, many game states are ambiguous and governed by convention, perceived kicker ability, and the expected quality of a team’s lineout. In contrast, American football has long relied on \emph{Expected Points} (EP), \emph{Expected Points Added} (EPA), and \emph{win-probability} (WP) frameworks to guide the analogous fourth-down decision, translating field position and context into expected returns and actionable recommendations \citep{Romer2006,Burke2009,YurkoVenturaHorowitz2018}. In that setting, WP and EP estimators are typically built using statistical machine learning models (e.g., random forests and gradient boosting), as in \cite{LockNettleton2014} and \cite{Baldwin2021}.

In rugby union, quantitative decision analysis has progressed more slowly, largely because detailed public data is limited. \cite{Martinez-Arastey2025} present a transparent phase-level Expected Points framework for the English Premiership and show the difficulty of predicting low-frequency outcomes such as penalty kicks. Separately, public analyses of goal-kicking estimate make probability from distance and angle, yielding useful but largely context-free benchmarks \citep{Monpezat2023}. Building on both strands---and on related decision-modeling work in American football---we evaluate the penalty decision directly.

This paper develops a decision-ready EP framework for the penalty decision. We estimate two context-aware surfaces: (i) expected points for possessions that begin with a lineout and (ii) expected points for penalty kicks at goal. The kick surface combines a continuous angle--distance make model with the \emph{continuation value} of misses (e.g., 22\,m drop-outs and in-play recoveries), rather than assigning misses zero value. Comparing these surfaces yields decision maps (zero-difference contours) that identify where each option maximizes expected return, and sensitivity analyses show how the frontier shifts with team quality, yellow/red cards, and expected meters gained to touch. The framework also makes explicit the translation from penalty location to resulting lineout location, along with the assumptions required when combining Premiership phase data with kicking data from Super Rugby, NPC (New Zealand Provincial Competition), and international matches.

\section{Data}

\subsection{Phase-Level Data}

Our primary dataset consists of phase-level event logs from the 2018/19 Premiership Rugby season compiled by \cite{Martinez-Arastey2025}, covering 132 matches and 35{,}199 phases of play. Each record describes the on-field state at the start of a phase: the initiating event type, a zonal field location and side, the current score differential, and the next score in points for or against the team in possession. We harmonize timestamps to ``seconds remaining in match'' so that time is strictly decreasing within a game and directly comparable across halves.

\begin{figure}[H]
    \centering
    \includegraphics[width=0.6\linewidth]{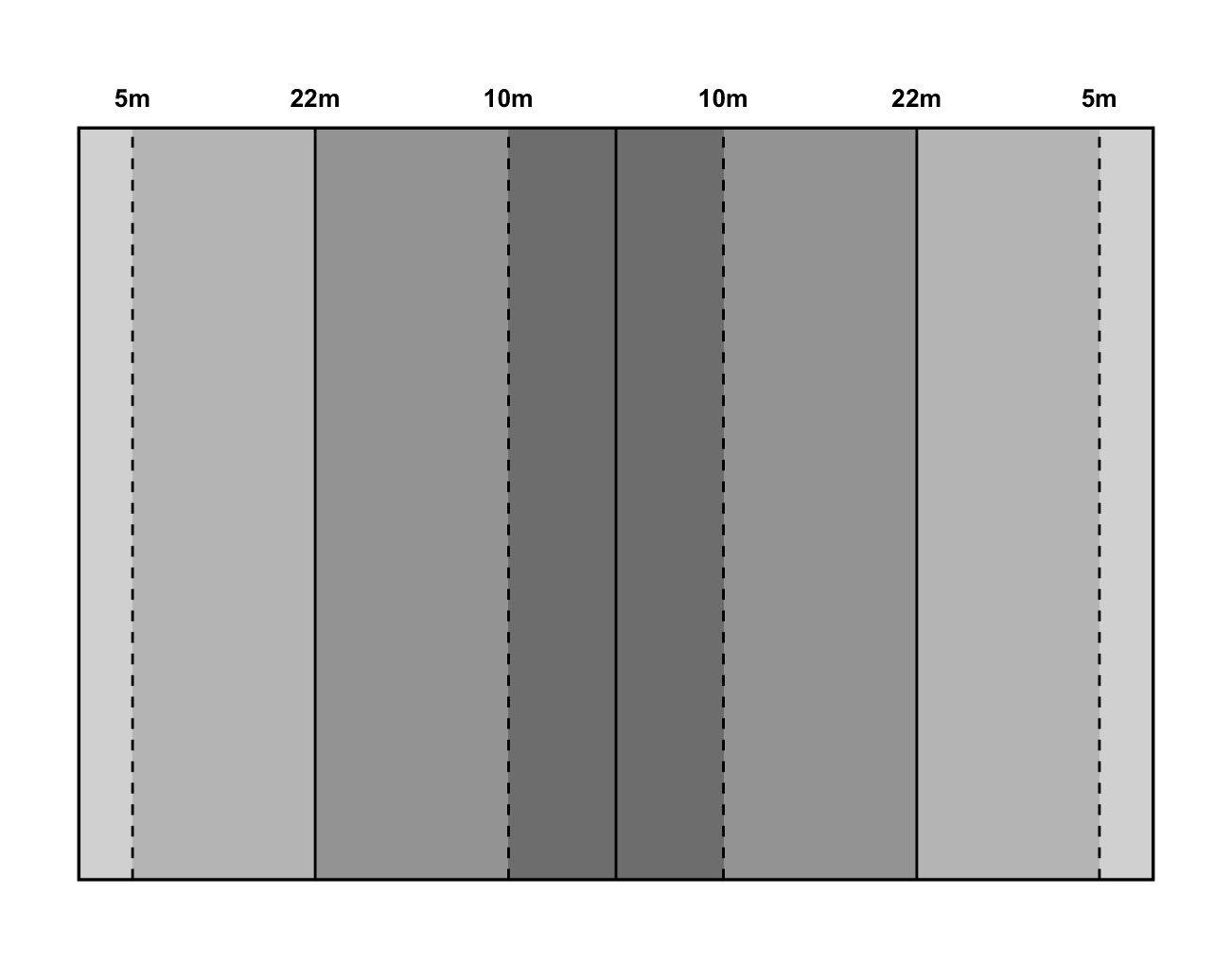}
    \caption{Field zones used in the phase-level dataset.}
    \label{fig:zones}
\end{figure}

The unit of observation in the dataset is a \emph{phase}, defined as the period between successive rucks. The phase count resets following restarts (e.g., lineouts, scrums, kick-offs) or major infringements that halt play. A \emph{possession sequence} is an ordered set of phases by the same team without an opposition touch that changes possession. Unless otherwise stated, Expected Points estimates condition on the \emph{first} phase of a possession sequence (e.g., a lineout), because that is the decision-relevant entry state. Figure~\ref{fig:zones} summarizes the zonal discretization of the pitch used in the phase-level data.

\subsection{Phase-Level Data Processing}

For the EP model, the unit of analysis is a possession that begins with a lineout. We therefore restrict the dataset to opening phases of play (i.e., observations with $\text{Phase} = 1$) whose initiating event is a lineout. Each such record represents a distinct possession entry state. We then construct several derived variables that encode field position, time, and game context:

\begin{itemize}[leftmargin=*]
    \item \textbf{Field position:} The phase data provide location as eight ordered zones (from opposition goal line to own goal line): \texttt{5m-Goal (opp)}, \texttt{22m-5m (opp)}, \texttt{10m-22m (opp)}, \texttt{Half-10m (opp)}, \texttt{10m-Half (own)}, \texttt{22m-10m (own)}, \texttt{5m-22m (own)}, \texttt{Goal-5m (own)}. We use this zonal location directly as the field-position factor in the lineout model.
    \item \textbf{Time remaining:} Total seconds remaining in the match are transformed to a half-specific variable, \texttt{Sec\_Remain\_Half}, defined as seconds remaining in the current half. We also define \texttt{Less\_Than\_2\_Min}, an indicator equal to 1 when \texttt{Sec\_Remain\_Half} $< 120$ and 0 otherwise.
    \item \textbf{Score and card context:} \texttt{Points\_Difference} records the score differential (for the team in possession), while \texttt{Card\_Diff} tracks the net player advantage in terms of yellow or red cards.
    \item \textbf{Team strength:} \texttt{WinPct\_Diff} is the running difference in win percentage between each team and its opponent at kickoff, with each team's running rate Laplace-smoothed toward 0.5 using $m$ artificial matches: $\widetilde{p}=(W+0.5m)/(G+m)$. We pre-specify $m=2$.
\end{itemize}

To address potential double counting across consecutive possessions, we create a grouping identifier, \texttt{run\_id}. Within each match, after ordering by event index, a run is defined as a maximal consecutive block of lineout-start possessions for the same team with unchanged starting point differential and unchanged eventual next-score outcome. Within each \texttt{run\_id} we compute \texttt{n\_same}, the run length. The role of \texttt{run\_id} in estimation is discussed in Section~\ref{sec:lineout}. The final structure of this reduced dataset is illustrated in Table~\ref{tab:reduced-data-structure}.

\begin{table}[htbp!]
\centering
\caption{A summary of the processed rugby union phase-level dataset.}
\label{tab:reduced-data-structure}
\resizebox{\textwidth}{!}{
\begin{tabular}{ccccccccccccccc}
\toprule
\textbf{ID} & \textbf{Round} & \textbf{Home} & \textbf{Away} & \textbf{Phase} &
\textbf{Team\_In\_Poss} & \textbf{Points\_Difference} & \textbf{Sec\_Remain\_Half} &
\textbf{Less\_Than\_2\_Min} & \textbf{Card\_Diff} & \textbf{WinPct\_Diff} & \textbf{Location} &
\textbf{run\_id} & \textbf{n\_same} & \textbf{Points} \\
\midrule
8  & 1 & Harlequins & Sale & 1 & Home &  0 & 2302 & 0 & 0 & 0 & 10m-22m (opp) & 1 & 1 &  3 \\
10 & 1 & Harlequins & Sale & 1 & Home &  3 & 2166 & 0 & 0 & 0 & Half-10m (opp) & 2 & 1 & -3 \\
14 & 1 & Harlequins & Sale & 1 & Away & -3 & 2103 & 0 & 0 & 0 & 22m-5m (opp) & 3 & 2 &  3 \\
17 & 1 & Harlequins & Sale & 1 & Away & -3 & 2043 & 0 & 0 & 0 & 10m-22m (opp) & 3 & 2 &  3 \\
53 & 1 & Harlequins & Sale & 1 & Home &  0 & 1313 & 0 & 0 & 0 & 10m-Half (own) & 4 & 1 &  7 \\
56 & 1 & Harlequins & Sale & 1 & Away &  0 & 1188 & 0 & 0 & 0 & 5m-22m (own) & 5 & 1 & -7 \\
67 & 1 & Harlequins & Sale & 1 & Away & -7 &  926 & 0 & 0 & 0 & 10m-22m (opp) & 6 & 1 &  7 \\
\bottomrule
\end{tabular}
}
\end{table}

\subsection{Kicking Data}

Our kicking dataset was generously provided by Ken Quarrie. It includes international penalty observations from \cite{QuarrieHopkins2015}, along with kicks from Super Rugby and the NPC competition. The sample spans 2000--2012 and contains 13,338 observations.

\begin{figure}[htbp!]
  \centering
  \includegraphics[width=0.75\linewidth]{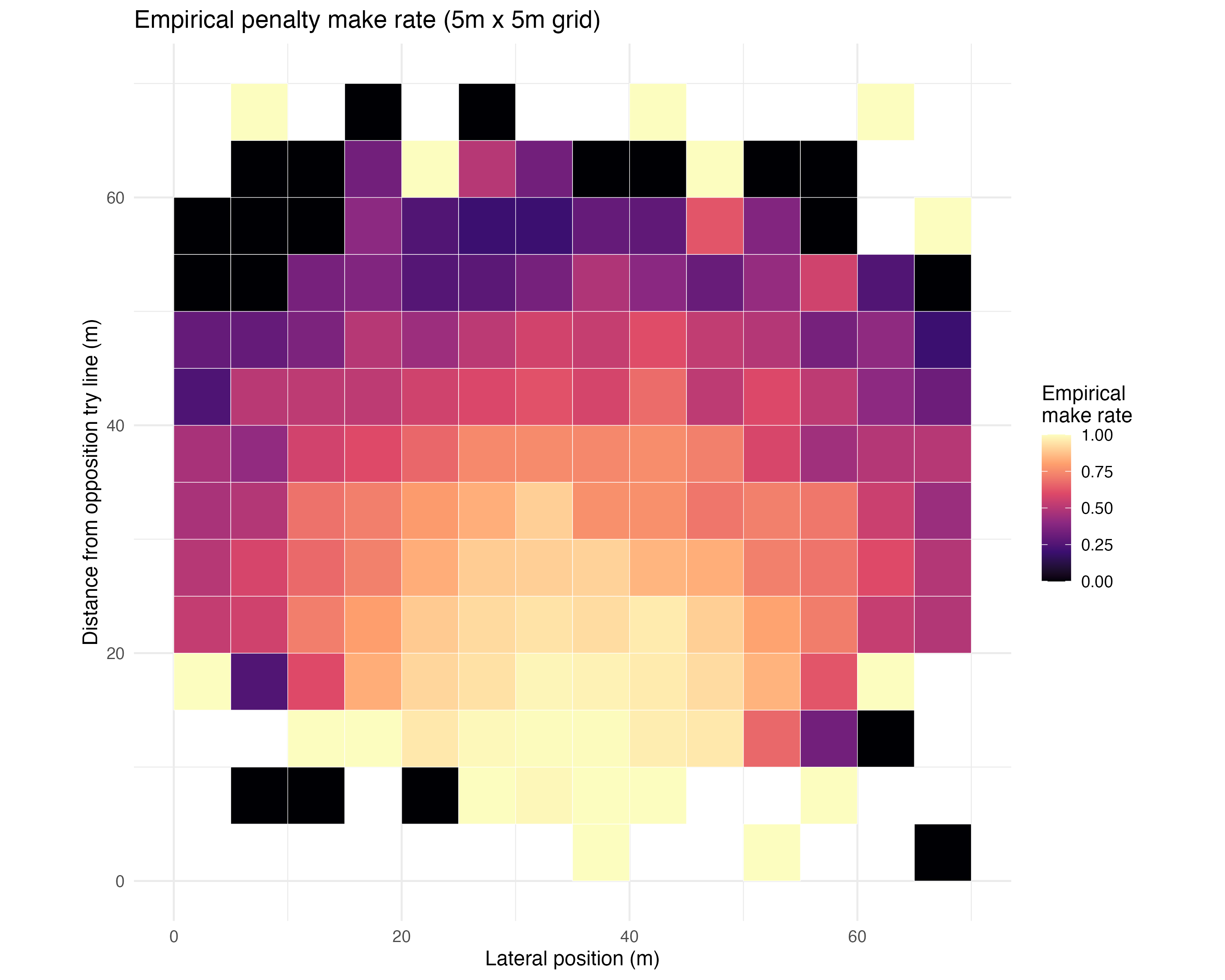}
  \caption{Empirical penalty-kick conversion rates by kick location, aggregated over 5\,m $\times$ 5\,m pitch bins.}
  \label{fig:kick-empirical-heatmap}
\end{figure}

Each record contains exact $(x, y)$ penalty locations, which lets us compute both kick distance and angle relative to the posts. Figure~\ref{fig:kick-empirical-heatmap} shows the empirical make rate in 5\,m $\times$ 5\,m pitch bins.

\section{Methods}

\subsection{Expected Points Framework}

We evaluate the penalty decision, kicking at goal versus kicking to touch, by comparing the expected points (EP) of the two options at a given field location. Let $(x, y)$ denote the location of a penalty on the field, where $x$ is the lateral distance in meters from the left touchline and $y$ is the distance in meters from the opposition try line. For a given assumed gain in territory from a kick to touch, denoted $d_{\text{touch}}$, we define the decision quantity as the difference in expected points, with positive values favoring a lineout:
\begin{equation}
\Delta EP(x, y; d_{\text{touch}}) = EP_{\text{lineout}}(y_{\text{LO}}) - EP_{\text{kick}}(x, y),
\label{eq:delta-ep}
\end{equation}
where $EP_{\text{kick}}(x, y)$ is the expected points from attempting a kick at goal from $(x, y)$, and $EP_{\text{lineout}}(y_{\text{LO}})$ is the expected points from a lineout at location $y_{\text{LO}}$ on the sideline. Here $y_{\text{LO}}$ denotes the resulting longitudinal coordinate after kicking to touch, with the translation rule defined in Section~\ref{sec:adjustment}.

The following subsections describe estimation of $EP_{\text{lineout}}$ and $EP_{\text{kick}}$.

\subsection{Expected Points of a Lineout}
\label{sec:lineout}

To value a lineout, we model the probability of each \emph{next} scoring outcome using a multinomial logistic regression, based on possession sequences that begin with a lineout in a given field zone. Let $\mathcal{C}$ denote the set of observed next-score outcomes in the sample (from the perspective of the team in possession), which in our data is $\{-7, -5, -3, 0, 3, 5, 7\}$. For each outcome $c \in \mathcal{C}$, the model estimates
\[
\log \frac{P(Y = c \mid \mathbf{X})}{P(Y = c^* \mid \mathbf{X})}
  = \alpha_c + \beta_{c,1} \cdot \text{Location}
  + \beta_{c,2} \cdot \text{Card\_Diff}
  + \beta_{c,3} \cdot \text{WinPct\_Diff}
  + \beta_{c,4} \cdot \text{L2M},
\]
where $c^*$ is the reference outcome category and \texttt{L2M} indicates fewer than two minutes remain in the half. The coefficients $\{\alpha_c, \beta_{c,1}, \beta_{c,2}, \beta_{c,3}, \beta_{c,4}\}$ are estimated via maximum likelihood. The field position predictor \texttt{Location} enters the model as a factor, allowing a flexible, non-linear relationship between field zone and scoring probability. Expected points are then recovered as a weighted sum over outcomes,
\[
EP_{\text{lineout}}(y) = \sum_{c \,\in\, \mathcal{C}} c \cdot \hat{P}(Y = c \mid \mathbf{X}),
\]
where $\hat{P}(Y = c \mid \mathbf{X})$ is the fitted probability from the multinomial model.

A multinomial model is preferable to a single linear regression on points because the response is a discrete state variable and the decision problem depends on the full conditional outcome distribution, not only its mean. Two game states can share similar mean points yet imply different risk profiles if probability mass is distributed differently across no-score, penalty, and try outcomes. Modeling $P(Y=c\mid\mathbf{X})$ directly preserves that structure, after which EP is obtained as a linear functional of estimated state probabilities, consistent with prior American-football EP modeling \citep{YurkoVenturaHorowitz2018}.

Grouped five-fold sensitivity checks by match (Appendix~\ref{app:model-sensitivity}, Table~\ref{tab:model-sensitivity}) support an additive baseline with \texttt{Location}, \texttt{Card\_Diff}, \texttt{WinPct\_Diff}, and \texttt{Less\_Than\_2\_Min}. We therefore keep \texttt{Less\_Than\_2\_Min} in the main specification so end-of-half effects enter the core model directly. Alternative covariate specifications (\texttt{Card\_Diff} $\times$ \texttt{WinPct\_Diff}, separate team-strength terms, and adding continuous \texttt{Sec\_Remain\_Half}) are reported in Appendix~\ref{app:model-sensitivity}. The win-percentage shrinkage prior is pre-specified at $m=2$ pseudo-games, with robustness checks reported in Appendix~\ref{app:winpct-prior-sensitivity}, Table~\ref{tab:winpct-prior-sensitivity}.

This specification captures downstream contingencies observed in matches: losing one's own lineout, turnovers in subsequent phases, or the opponent scoring following a turnover. The resulting quantity represents the \emph{expected next-score value} of having a lineout at that field position, given competition-level play and the current game context. For interpolation and plotting, we order zones from opposition to own goal line and map them to zone midpoints in meters from the opposition try line $(5,\ 13.5,\ 31,\ 45,\ 55,\ 69,\ 86.5,\ 97.5)$ using boundaries $(0,\ 5,\ 22,\ 40,\ 50,\ 60,\ 78,\ 95,\ 100)$. Unless stated otherwise, decision maps set \texttt{Less\_Than\_2\_Min} to 0.

\subsubsection{Dealing with Double Counting}

Double counting can arise not only at the phase level (which we already handle by restricting to $\text{Phase} = 1$), but also across consecutive possessions that map to the same next-score event. Table~\ref{tab:example-data} shows three successive lineout-start possessions for the same team with the same starting score differential and the same eventual outcome.

\begin{table}[htbp!]
\centering
\caption{An example of consecutive lineout-start possessions mapped to the same next-score outcome.}
\label{tab:example-data}
\resizebox{\textwidth}{!}{
\begin{tabular}{ccccccccccccccc}
\toprule
\textbf{ID} & \textbf{Round} & \textbf{Home} & \textbf{Away} & \textbf{Phase} & \textbf{Team\_In\_Poss} & \textbf{Points\_Difference} & \textbf{Sec\_Remain\_Half} & \textbf{Less\_Than\_2\_Min} & \textbf{Card\_Diff} & \textbf{WinPct\_Diff} & \textbf{Location} & \textbf{run\_id} & \textbf{n\_same} & \textbf{Points} \\
\midrule
19 & 1 & Harlequins & Sale & 1 & Home & 21 & 454 & 0 & 0 & 0 & 22m-10m (own) & 11 & 3 & 7 \\
20 & 1 & Harlequins & Sale & 1 & Home & 21 & 378 & 0 & 0 & 0 & 22m-5m (opp) & 11 & 3 & 7 \\
21 & 1 & Harlequins & Sale & 1 & Home & 21 & 146 & 0 & 0 & 0 & 22m-5m (opp) & 11 & 3 & 7 \\
\bottomrule
\end{tabular}
}
\end{table}

Let run $r$ contain $n_r$ consecutive lineouts with a common eventual next-score outcome $Y_r$. If all rows are kept, the likelihood receives $n_r$ copies of $Y_r$ from that run, so long runs get mechanically up-weighted. In effect, a single scoring event can be counted multiple times simply because play recycled through repeated lineouts before that score occurred. This would bias fitted class probabilities toward outcomes that happen to appear in long same-outcome runs, rather than toward genuinely more frequent scoring states.

Accordingly, we retain exactly one representative row per \texttt{run\_id}. For run $r$ with observations $Y_{r1},\dots,Y_{rn_r}$, we draw one index uniformly and keep only that row:
\[
J_r \sim \text{Uniform}\big(\{1, \dots, n_r\}\big).
\]
We implement this directly through row selection (one retained row per run) rather than through a weighted-likelihood fit.

This run-level subsampling ensures that each repeated same-outcome sequence contributes once, mitigating pseudo-replication while preserving realistic variation in field position and context across runs. The same idea is used inside each bootstrap replicate (one draw per run in each resampled match), so point estimates and uncertainty intervals are aligned to the same dependence structure. Our approach is conceptually similar to the cluster-level subsampling used by \citet{BrillYurkoWyner2025}, who randomly retain a subset of plays per simulated game to reduce within-game dependence when estimating win probability. After sampling, we obtain a total of 2{,}046 lineout observations for estimation.

\subsubsection{Multinomial Regression Summary}

\begin{figure}[htbp!]
    \centering
    \includegraphics[width=0.75\linewidth]{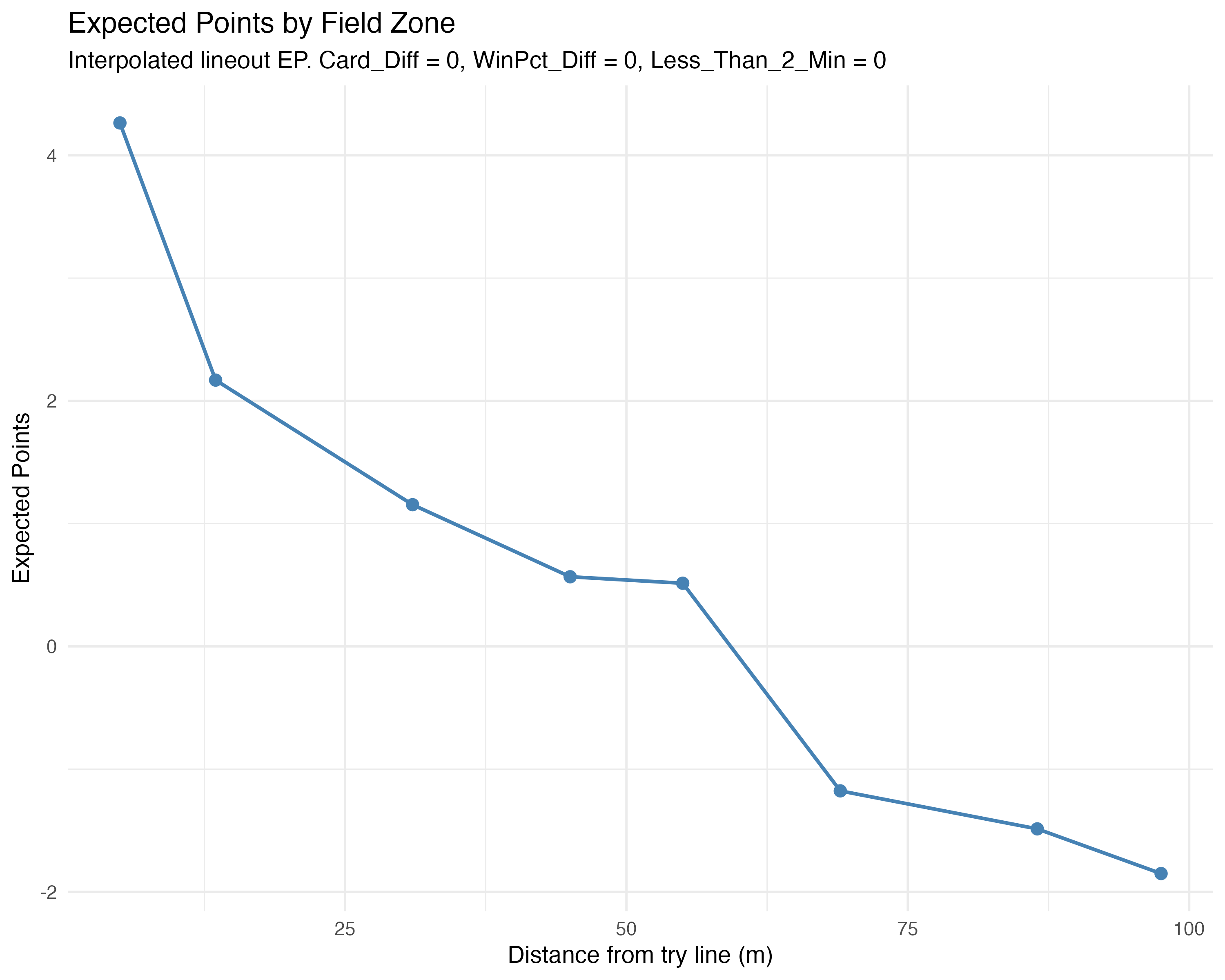}
    \caption{Estimated expected points of the next score by lineout meter line (interpolated from zonal field-position factors).}
    \label{fig:ep-lineout}
\end{figure}

The lineout model treats the next scoring event as a categorical random variable with outcomes $\{-7, -5, -3, 0, 3, 5, 7\}$ points and converts fitted class probabilities into EP through the probability-weighted sum defined above (Figure~\ref{fig:ep-lineout}).

\subsubsection{EP Uncertainty Quantification via Bootstrap}

\begin{figure}[htbp!]
    \centering
    \includegraphics[width=0.75\linewidth]{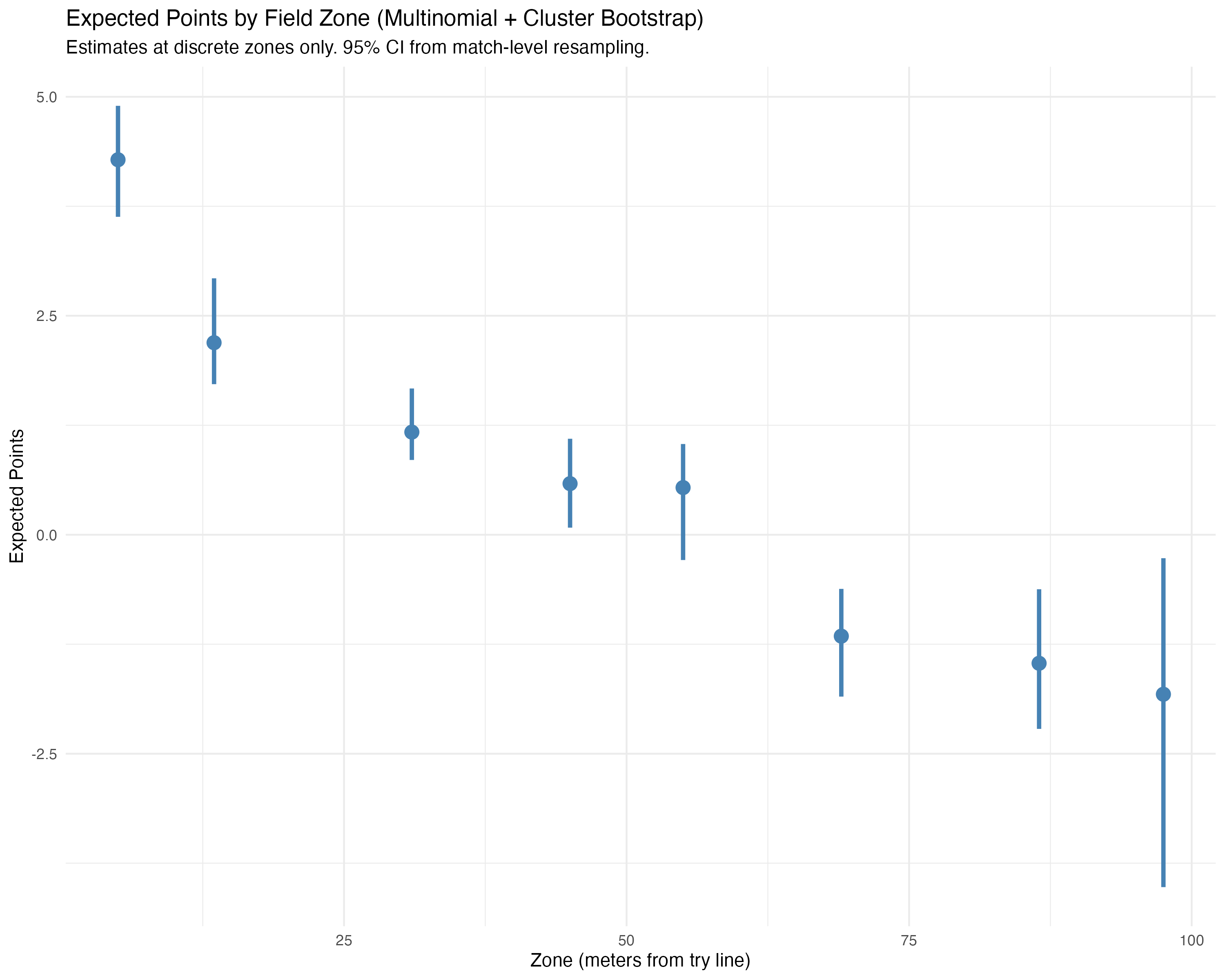}
    \caption{Bootstrap uncertainty bands for lineout expected points (95\% confidence intervals).}
    \label{fig:ep-lineout-bootstrap}
\end{figure}

To quantify uncertainty in the expected points estimates, we use a percentile bootstrap with $B = 2000$ replicates. Each replicate follows the same dependence-aware procedure:
\begin{enumerate}[leftmargin=*]
  \item Sample matches with replacement (match-level cluster bootstrap).
  \item Within each resampled match, and for each \texttt{run\_id}, draw exactly one row uniformly at random.
  \item Refit the multinomial model on that sampled dataset.
  \item Recompute lineout EP across meter-line positions from fitted class probabilities.
\end{enumerate}
Confidence intervals are given by the 2.5th and 97.5th percentiles of the bootstrap distribution (Figure~\ref{fig:ep-lineout-bootstrap}).

Uncertainty is largest near a team's own try line, where lineout-start possessions are sparse and penalty shots at goal wouldn't even be considered. The decision analysis therefore focuses on penalties within roughly 65\,m of the opposition try line.

\subsection{Expected Points of a Penalty Kick} \label{sec:kick-at-goal}

The expected points of a penalty kick at goal are the weighted sum of the value of a make and the continuation value of a miss. Unlike lineouts, kick outcomes depend primarily on distance and angle to the posts rather than raw field coordinates. Let $d$ denote distance from the penalty spot to the center of the posts and let $\theta$ denote the lateral angle. Then
\begin{equation}
EP_{\text{kick}}(d, \theta) =
\P_{\text{make}}(d, \theta) \cdot 3
+ \big(1 - \P_{\text{make}}(d, \theta)\big) \cdot EP_{\text{miss}}(d, \theta),
\label{eq:ep-kick}
\end{equation}
where 3 points are awarded for a successful kick, $\P_{\text{make}}(d, \theta)$ is the probability of a successful kick from $(d, \theta)$, and $EP_{\text{miss}}(d, \theta)$ is the expected points of the continuation state following a miss (e.g., change in possession, change in field position).

The field coordinates $(x, y)$ of a penalty uniquely determine $(d, \theta)$ through pitch geometry, so $EP_{\text{kick}}(x, y)$ is obtained by evaluating $EP_{\text{kick}}(d, \theta)$ at the corresponding $(d, \theta)$. We estimate $\P_{\text{make}}(d, \theta)$ from the kicking data and approximate $EP_{\text{miss}}(d, \theta)$ from restart sequences in the phase-level dataset.

\subsection{Estimating Kick Success Probability}

Using the exact $(x, y)$ penalty locations, we compute the kick angle to the posts as:
\[
\theta = \arctan\left(\frac{|x - 35|}{y}\right) \cdot \frac{180}{\pi}
\]
where $x$ is the distance from the left touchline, $y$ is the distance from the opposition try line, and 35 is the midpoint of the pitch width (in meters), so $|x - 35|$ is the horizontal distance from the center of the posts (Figure~\ref{fig:angle}).

\begin{figure}[htbp!]
\centering
\begin{tikzpicture}[scale=1.2]

  \draw[thick] (0, 0) -- (6, 0) node[right] {try line};

  \filldraw[black] (3, 0) circle (2pt) node[below] {posts $(35, 0)$};

  \filldraw[black] (4.8, 3) circle (2pt) node[right] {kick $(x, y)$};

  \draw[dashed] (3, 0) -- (3, 3.4);

  \draw[thick] (3, 0) -- (4.8, 3);

  \draw[->] (3, 0.8) arc (90:58:0.8) node[midway, above right] {$\theta$};

  \draw[<->] (3, 3.3) -- (4.8, 3.3) node[midway, above] {$|x - 35|$};

  \draw[<->] (2.8, 0) -- (2.8, 3) node[midway, left] {$y$};

  \draw[thick] (3, 0) -- (4.8, 3) node[midway, right] {distance};

\end{tikzpicture}
\caption{Geometry used to compute the kick angle $\theta$ from field coordinates $(x, y)$, where $y$ is the distance from the try line and $|x - 35|$ is the horizontal distance from the center of the posts.}
\label{fig:angle}
\end{figure}

We then fit a logistic model with distance from the posts and angle as the two predictors:
\[
\log\left(\frac{p_i}{1 - p_i}\right) = \beta_0 + \beta_1 \cdot \text{distance}_i + \beta_2 \cdot \theta_i
\]
where $p_i$ is the probability that kick $i$ is successful.

As a sensitivity check (Appendix~\ref{app:model-sensitivity}, Table~\ref{tab:model-sensitivity}), a GAM with smooth terms in angle and distance produced nearly identical out-of-sample performance (log loss 0.5164 vs.\ 0.5184 for logistic) and similar marker-level make probability in the case study (0.563 vs.\ 0.538). We therefore retain the logistic specification for interpretability.

\begin{figure}[H]
    \centering
    \includegraphics[width=0.68\linewidth]{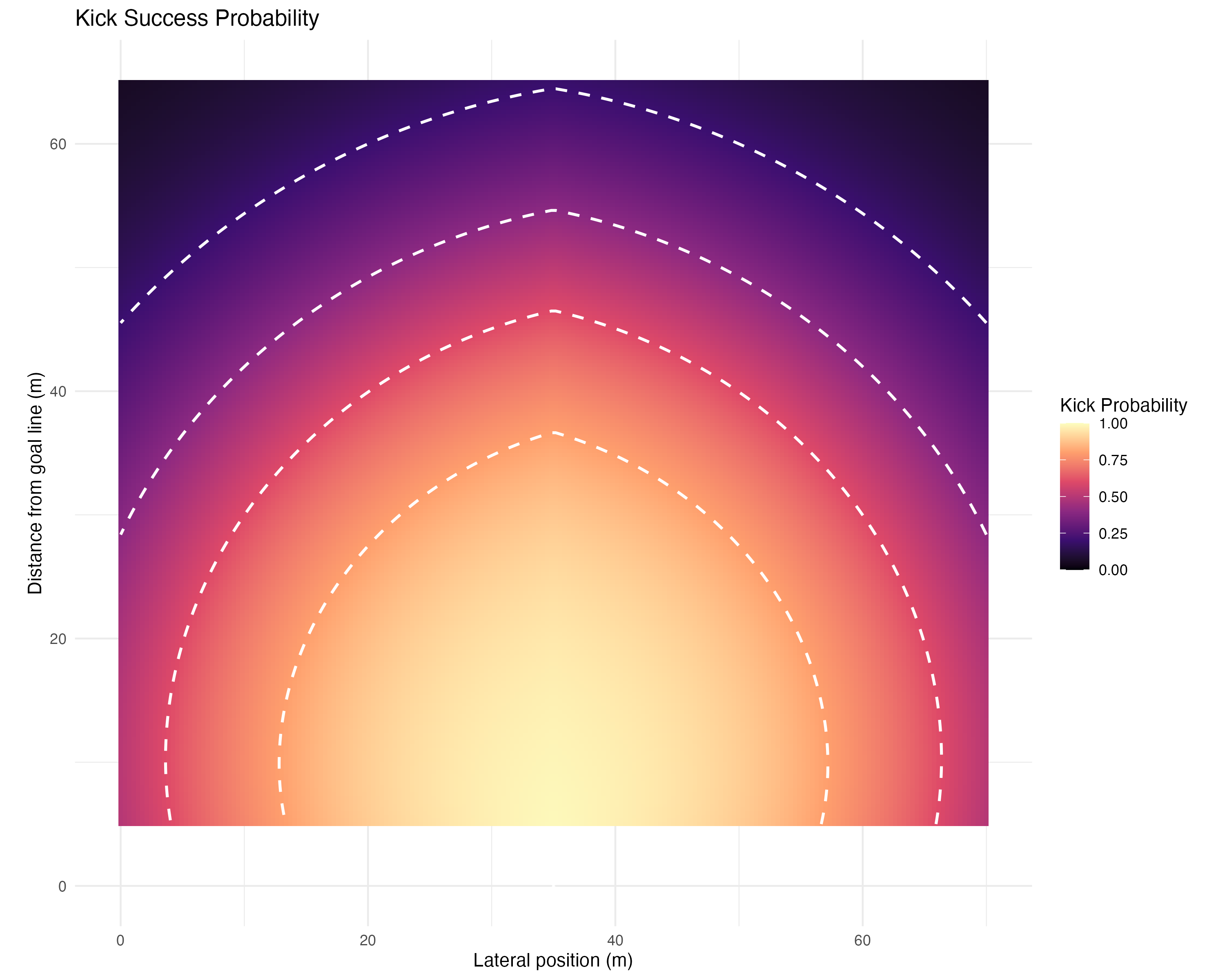}
    \caption{Estimated penalty-kick success probabilities by field position. The goal-post center is at lateral position 35\,m; estimates begin 5\,m from the goal line because no kicks occur at 0\,m.}
    \label{fig:kick-success-prob}
\end{figure}

Figure~\ref{fig:kick-success-prob} shows the estimated kick success probabilities, with dashed contours at probability levels $0.2$, $0.4$, $0.6$, and $0.8$.

\subsection{Continuation Value of a Missed Kick}

In rugby, a missed penalty does not immediately end the sequence. After a miss, the defending team may ground the ball in-goal, the ball may travel dead (triggering a 22\,m drop-out), or the ball may remain live and be returned from inside the 22. Returned-in-play misses are relatively infrequent, and our phase labels do not reliably separate them from other open-play kicks. We therefore adopt a simplifying assumption: a missed penalty transitions to a 22\,m drop-out restart.

Operationally, we identify restarts that (i) do not follow a change in score (excluding post-score kick-offs) and (ii) did not occur at the start of a half, and compute the expected next-score value of those entries by restart zone, using the same EP framework as for lineouts (Section~\ref{sec:lineout}). Table~\ref{tab:ep-location} reports average expected points following kick restarts by original kick location. Because restart location is only available in broad zones, we map each zone to its midpoint in meters from the opposition try line and fit a weighted smoothing spline (weights = zone counts) to obtain a continuous approximation for $EP_{\text{miss}}(y)$. This avoids artificial step changes at zone boundaries while remaining explicit about the coarse zonal input.

\begin{table}[htbp!]
\centering
\caption{Average points following missed-kick restarts by phase-data zone.}
\label{tab:ep-location}
\begin{tabular}{lcc}
\toprule
\textbf{Location} & \textbf{n} & \textbf{Avg. Points} \\
\midrule
Goal--5m (own)  & 14 &  0.36 \\
5m--22m (own)   & 21 & -1.00 \\
22m--10m (own)  & 89 &  1.10 \\
10m--Half (own) & 45 & -0.44 \\
Half--10m (opp) & 22 &  1.23 \\
10m--22m (opp)  & 7  &  4.29 \\
\midrule
\textbf{Overall Average} & 198 & 0.60 \\
\bottomrule
\end{tabular}
\end{table}

\subsection{Penalty Kick Expected Points Surface}

Combining the kick success surface and the continuation values yields an expected points surface for penalty kicks. For each location on the field, we map $(x, y)$ to $(d, \theta)$, evaluate $\P_{\text{make}}(d, \theta)$ from the logistic model, and evaluate smoothed $EP_{\text{miss}}(y)$ from the zonal restart data in Table~\ref{tab:ep-location}. These quantities are substituted into Equation~\eqref{eq:ep-kick} to obtain the expected points of a kick at goal from that location.

\subsection{Uncertainty Quantification for Kicks}

\begin{figure}[H]
    \centering
    \includegraphics[width=0.75\linewidth]{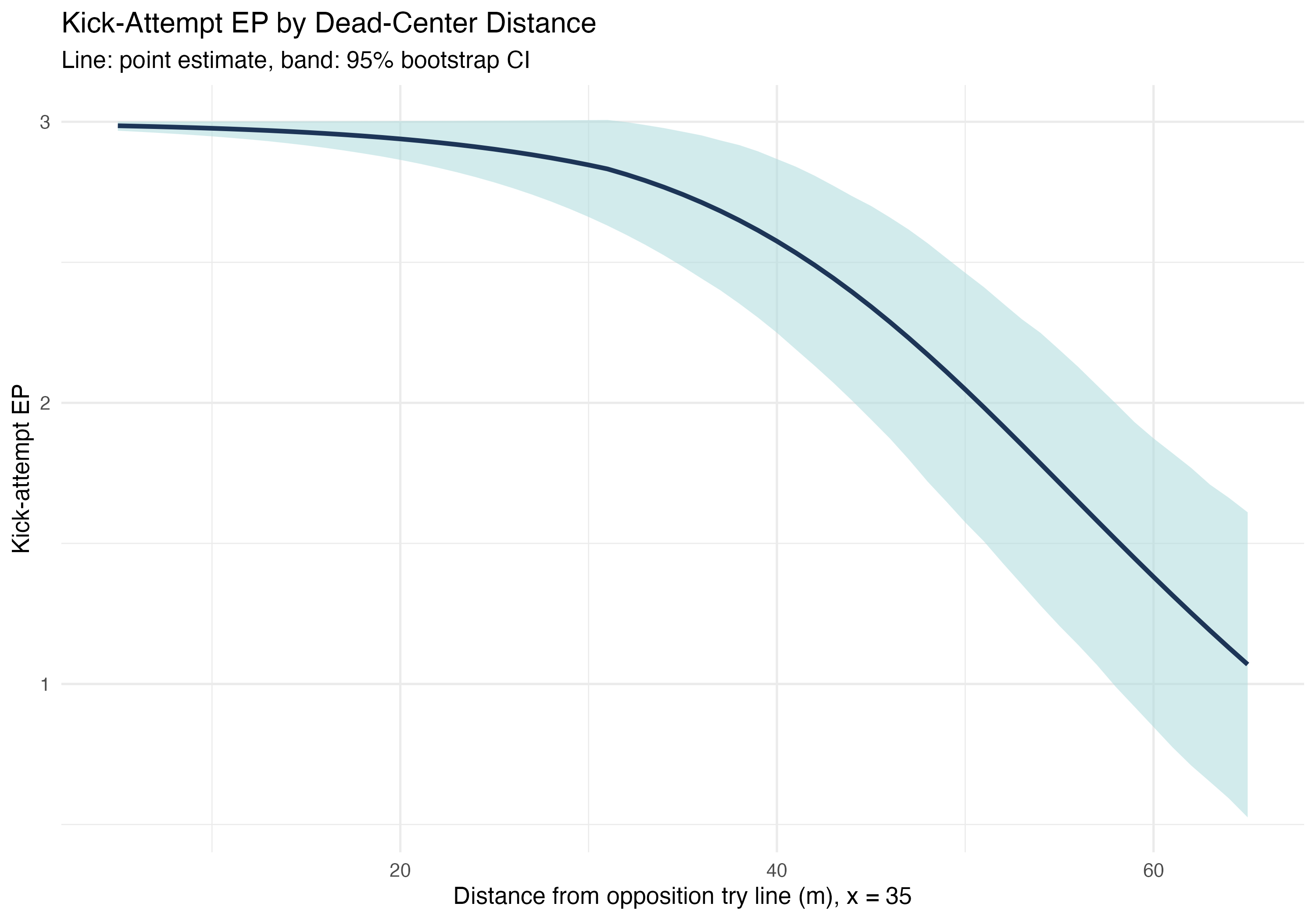}
    \caption{Expected points for dead-center kick attempts ($x=35$) as a function of distance from the goal line. The curve shows the bootstrap mean with a 95\% confidence band.}
    \label{fig:kick-ep-bootstrap}
\end{figure}

To align uncertainty reporting with the decision quantities of interest, we estimate uncertainty directly for kick-attempt expected points $EP_{\text{kick}}$ rather than reporting continuation-only and make-probability components separately. We use a bootstrap with $B = 2000$ replicates and summarize kick-attempt EP across dead-center kick locations ($x=35$) as distance from the opposition try line varies. In each replicate we resample penalty kicks and refit the kick-success model, resample restart rows and rebuild the continuation smoother, and then recompute $EP_{\text{kick}}$ over the dead-center distance grid (Figure~\ref{fig:kick-ep-bootstrap}).

At the case-study marker $(x=16, y=40)$ with $d_{\text{touch}}=20$\,m, this gives $EP_{\text{kick}}=2.22$ with 95\% CI $[1.65,2.76]$.

To propagate uncertainty into the \emph{decision} quantity itself, we run a joint bootstrap with $B = 2000$ replicates. In each replicate, we:
\begin{enumerate}[leftmargin=*]
  \item Resample matches with replacement and, within each resampled match, sample one lineout row per \texttt{run\_id}.
  \item Refit the lineout multinomial model and lineout smoother.
  \item Resample penalty kicks and refit the kick-success model.
  \item Resample restart rows, refit the continuation smoother, and recompute $EP_{\text{kick}}$ and $\Delta EP$.
\end{enumerate}
At the case-study marker $(x=16, y=40)$ with $d_{\text{touch}}=20$\,m, this yields $EP_{\text{lineout}}=1.69$ (95\% CI: $[1.25,2.20]$), $EP_{\text{kick}}=2.22$ (95\% CI: $[1.65,2.76]$), and $\Delta EP=-0.53$ (95\% CI: $[-1.26,0.25]$). The corresponding decision uncertainty is shown in Section~\ref{sec:single-penalty}.

\subsection{Adjusting for Meters Gained on Kick to Touch} \label{sec:adjustment}

At each penalty location on the field, we compare the value of attempting a kick at goal with the value of kicking to touch and taking the ensuing lineout. The kick-at-goal value is calculated at the penalty spot, using the distance and lateral angle to the posts implied by that location. By contrast, the lineout value is calculated at the location where the lineout would actually occur \emph{after} the kick to touch.

Let $(x, y)$ denote the penalty location, and let $d_{\text{touch}}$ denote the expected meters gained from a kick to touch along the length of the field. Starting from $y$ (distance from the opposition try line), we approximate the resulting lineout location as
\begin{equation}
y_{\text{LO}} = \max\{5,\; y - d_{\text{touch}}\},
\label{eq:y-lo}
\end{equation}
shifting the ball $d_{\text{touch}}$ meters closer to the opposition try line and truncating at the 5\,m line if necessary, since the laws prohibit lineouts closer than 5\,m from the try line. In reality, the ball is kicked into touch and the restart is on the sideline; however, our lineout EP model depends on longitudinal location ($y_{\text{LO}}$) and context covariates, not on lateral coordinate, so we suppress the lateral dimension for the lineout component.

For a given assumed gain $d_{\text{touch}}$, the expected points of choosing the lineout option from penalty location $(x, y)$ are then
\[
EP_{\text{lineout}}(y_{\text{LO}}),
\]
evaluated via the regression in Section~\ref{sec:lineout}. In our baseline decision maps we consider a grid of plausible translation distances
\[
d_{\text{touch}} \in \{0, 5, 10, 15, 20, 25\}\,\text{m},
\]
reflecting different assumptions about territory gained by kicking to touch. Because the data do not record realized meters gained to touch at the penalty level, $d_{\text{touch}}$ is treated as a scenario parameter rather than a latent random variable. The maps are therefore conditional on the chosen $d_{\text{touch}}$ value.

\subsection{Decision Maps}

Given $EP_{\text{lineout}}$ and $EP_{\text{kick}}$, the relative value of a lineout versus a kick at goal at the same penalty location is defined by Equation~\eqref{eq:delta-ep}. Here $(x, y)$ is the penalty location, $d_{\text{touch}}$ is the assumed meters gained by kicking to touch, and $y_{\text{LO}}$ is set by Equation~\eqref{eq:y-lo}. Positive values of $\Delta EP(x, y; d_{\text{touch}})$ favor choosing the lineout, while negative values favor taking the kick at goal, with $\Delta EP(x, y; d_{\text{touch}}) = 0$ tracing out the indifference frontier between the two options.

\begin{figure}[htbp!]
  \centering
  \begin{subfigure}[c]{0.32\linewidth}
    \includegraphics[width=\linewidth]{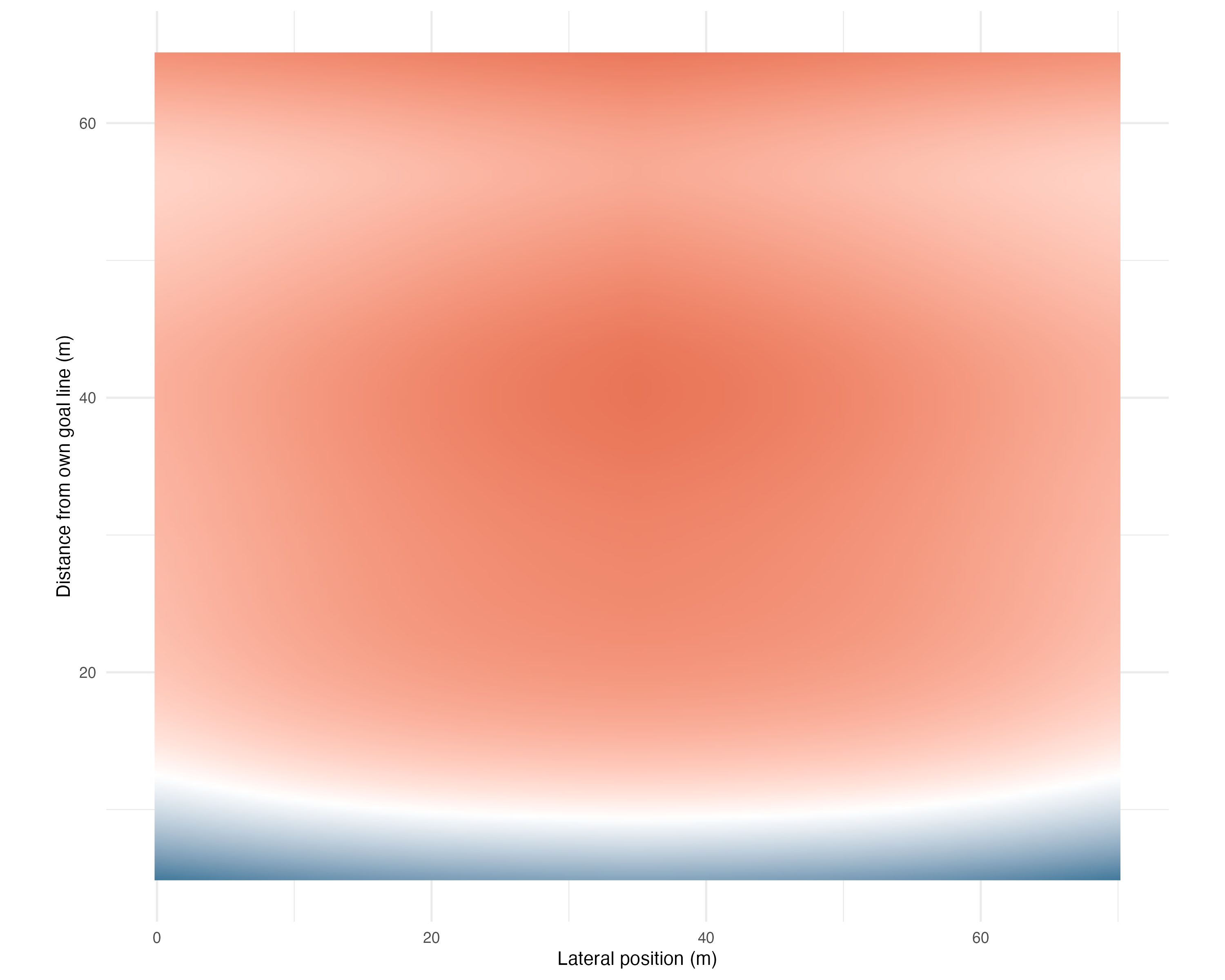}
    \caption{$d_{\text{touch}} = 0$\,m.}
    \label{fig:kick_lineout_0}
  \end{subfigure}
  \hfill
  \begin{subfigure}[c]{0.32\linewidth}
    \includegraphics[width=\linewidth]{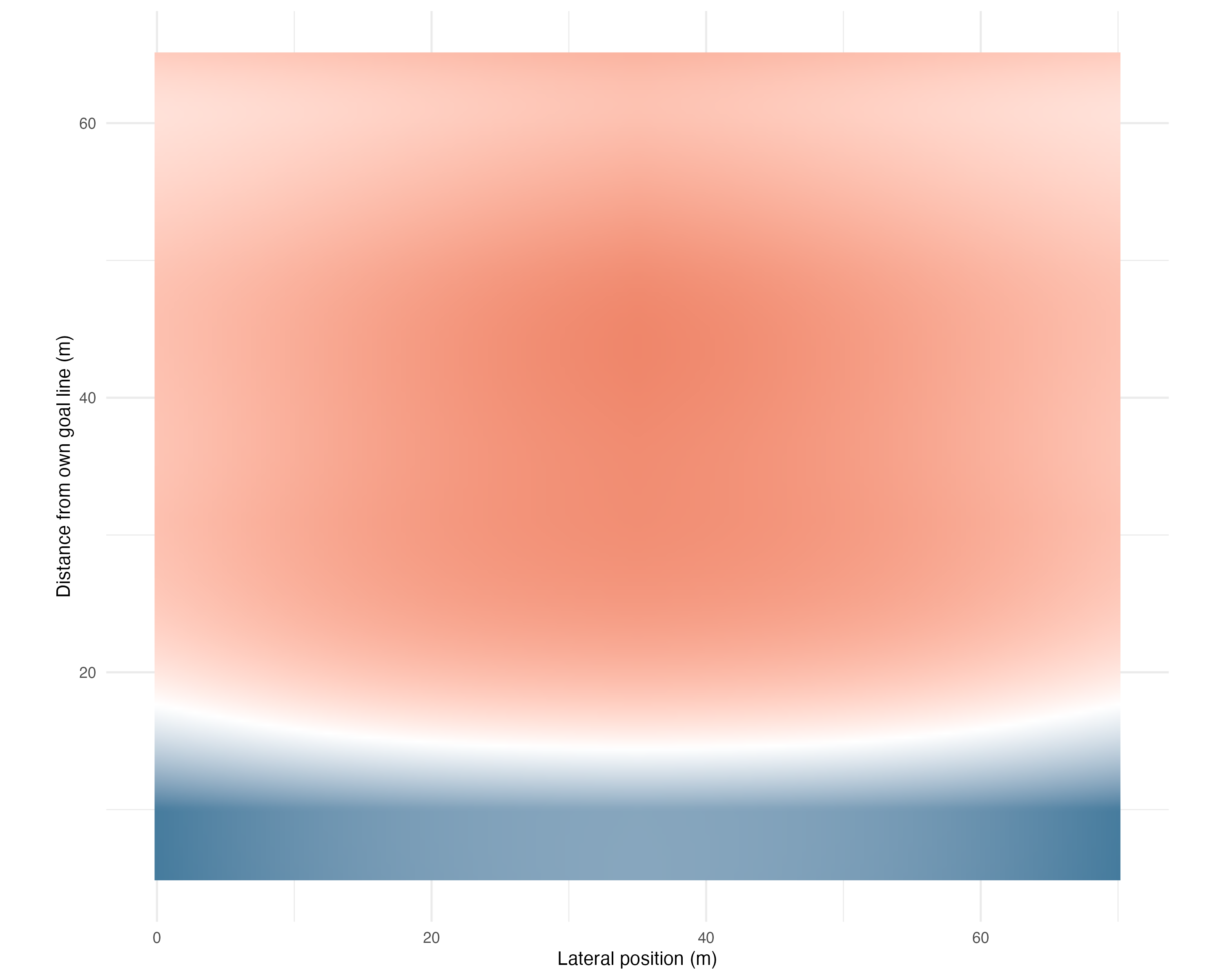}
    \caption{$d_{\text{touch}} = 5$\,m.}
    \label{fig:kick_lineout_5}
  \end{subfigure}
  \hfill
  \begin{subfigure}[c]{0.32\linewidth}
    \includegraphics[width=\linewidth]{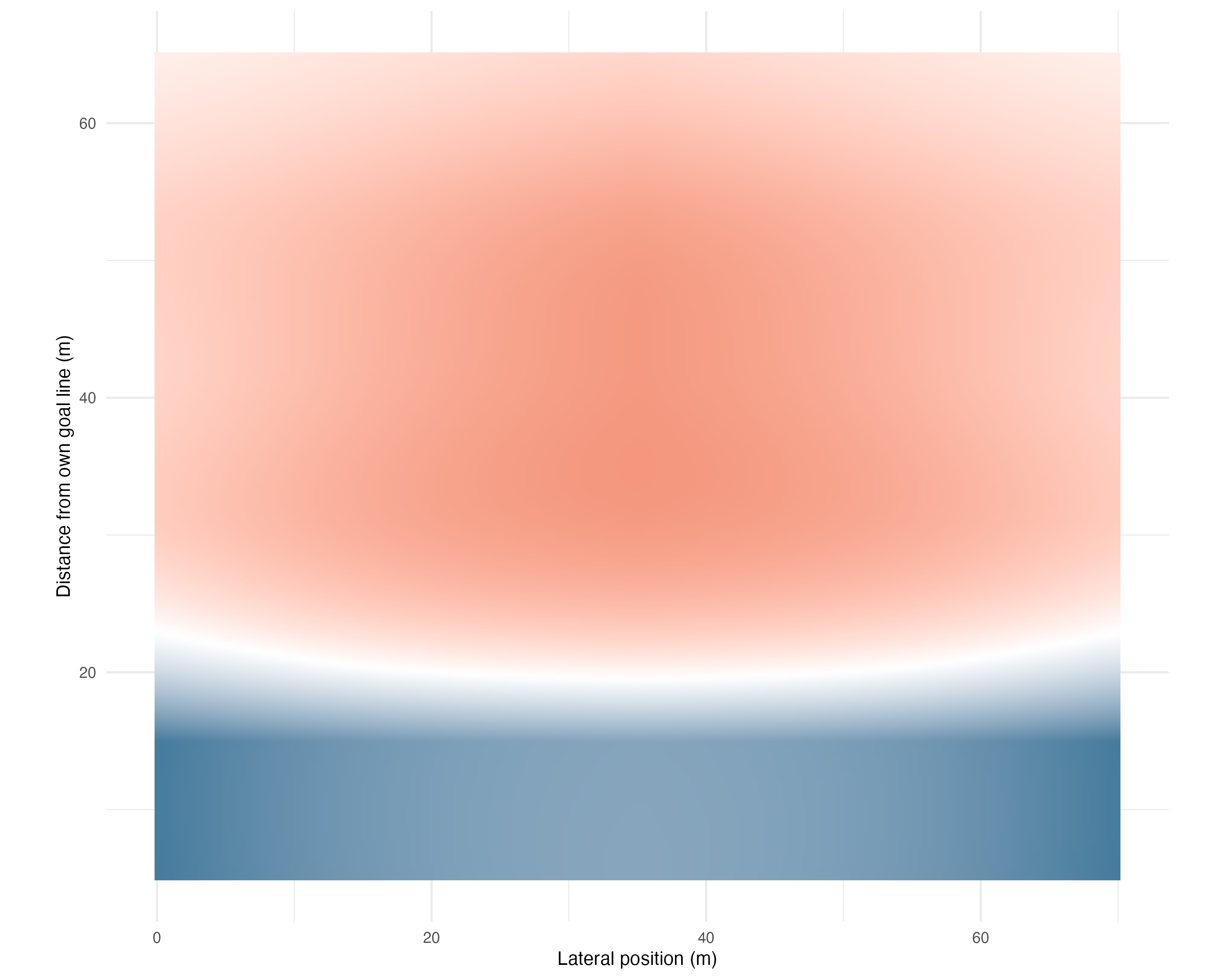}
    \caption{$d_{\text{touch}} = 10$\,m.}
    \label{fig:kick_lineout_10}
  \end{subfigure}
  \vspace{0.35em}
  \begin{subfigure}[c]{0.32\linewidth}
    \includegraphics[width=\linewidth]{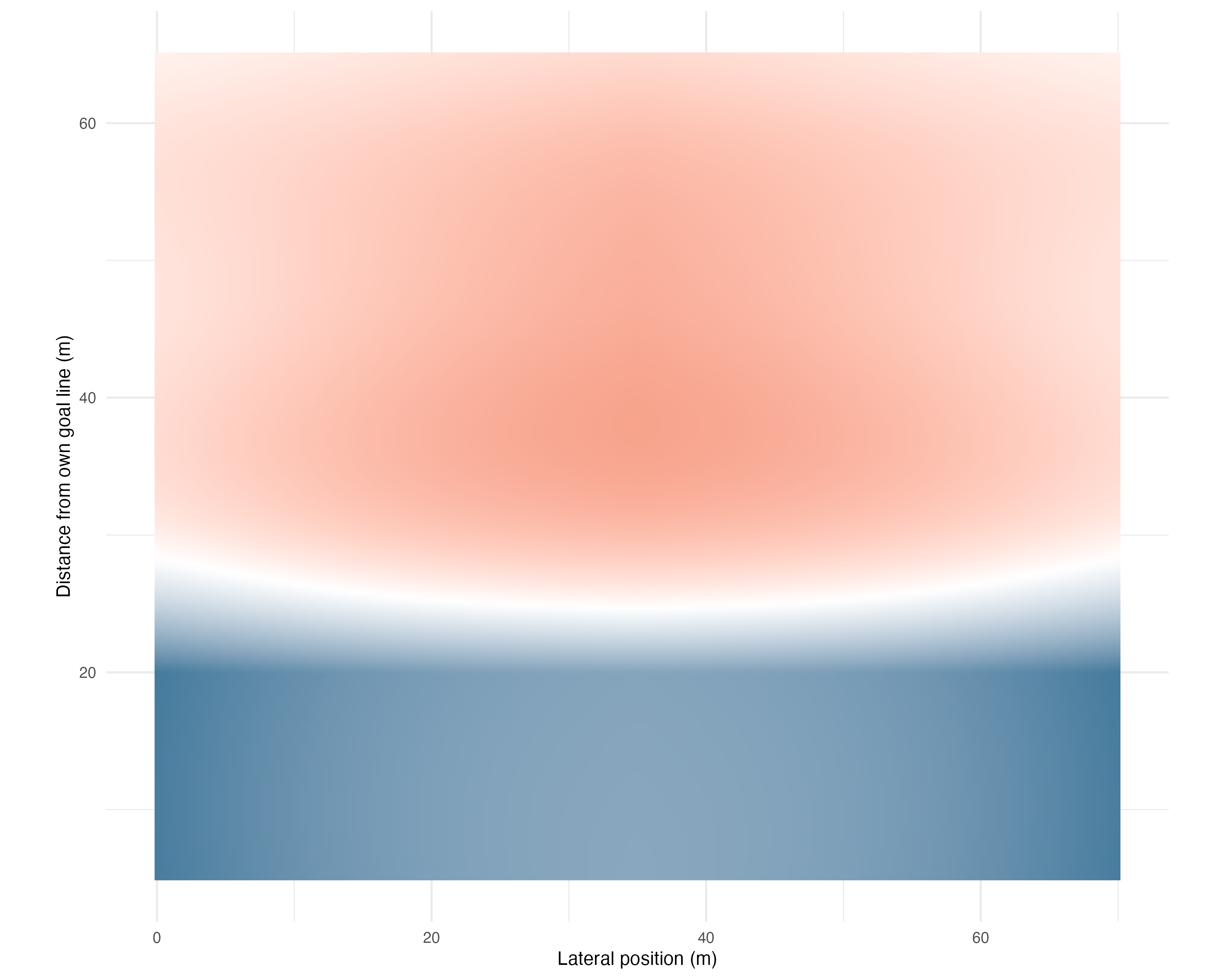}
    \caption{$d_{\text{touch}} = 15$\,m.}
    \label{fig:kick_lineout_15}
  \end{subfigure}
  \hfill
  \begin{subfigure}[c]{0.32\linewidth}
    \includegraphics[width=\linewidth]{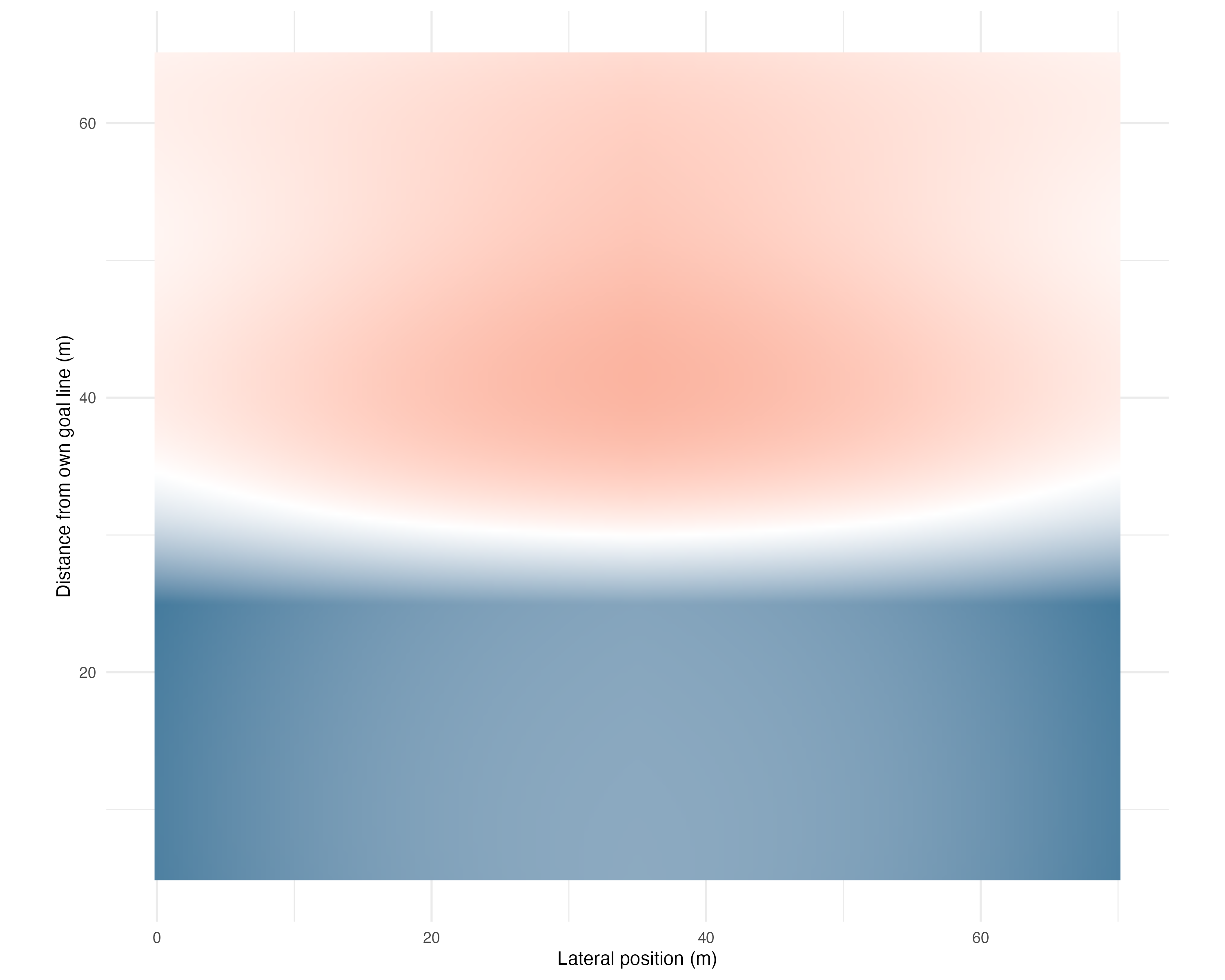}
    \caption{$d_{\text{touch}} = 20$\,m.}
    \label{fig:kick_lineout_20}
  \end{subfigure}
  \hfill
  \begin{subfigure}[c]{0.32\linewidth}
    \includegraphics[width=\linewidth]{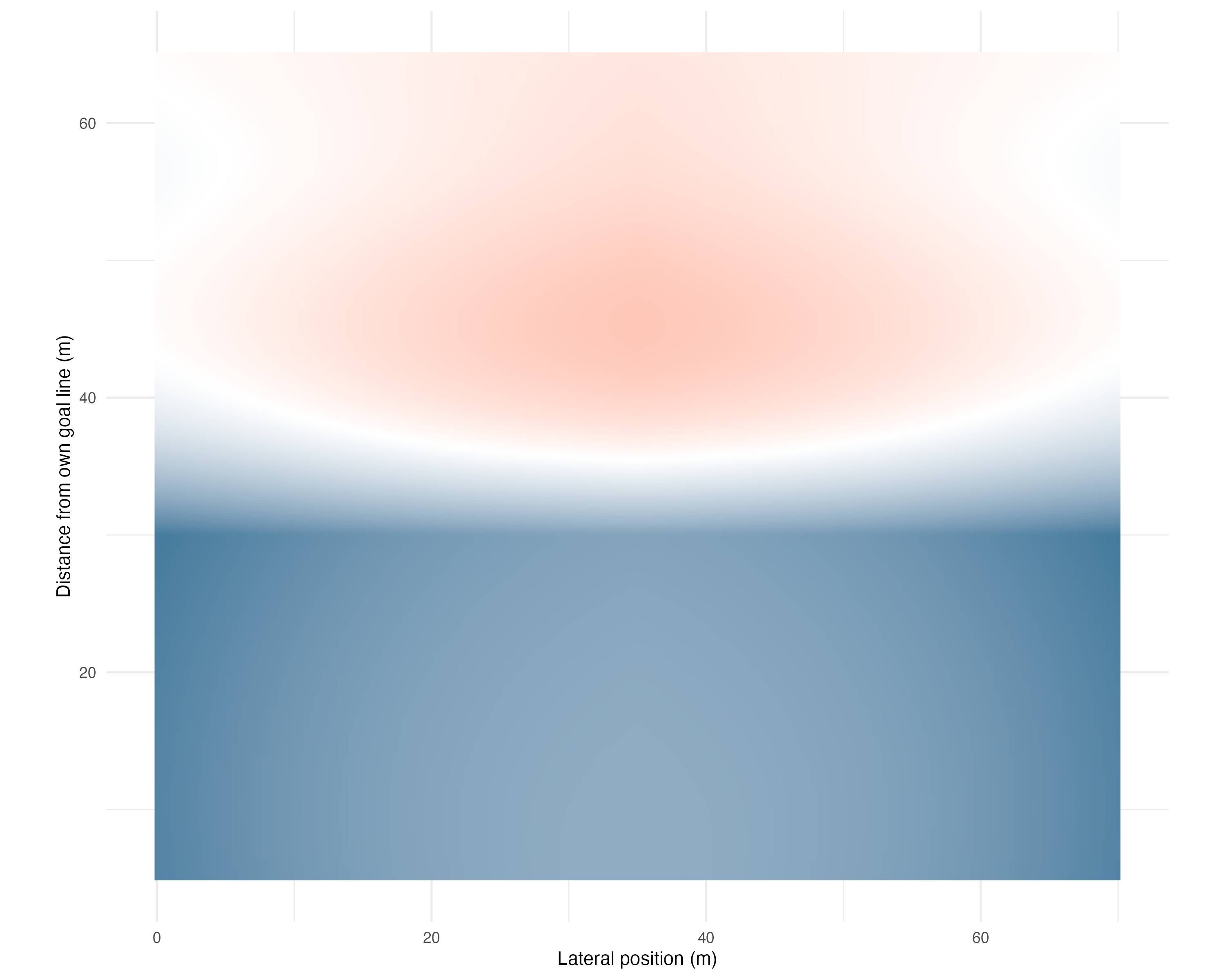}
    \caption{$d_{\text{touch}} = 25$\,m.}
    \label{fig:kick_lineout_25}
    \end{subfigure}
  \begin{subfigure}[c]{\linewidth}
    \centering
    \includegraphics[width=0.6\linewidth]{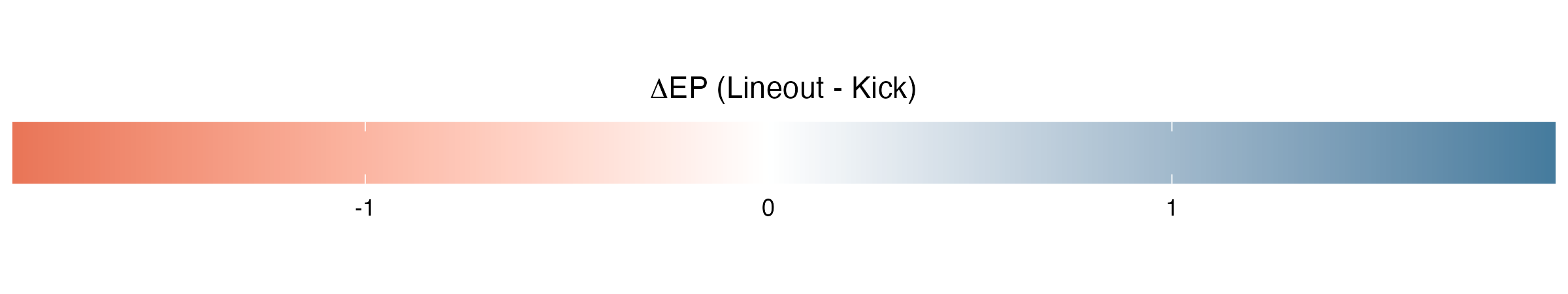}
  \end{subfigure}
  \caption{Decision maps comparing the expected points from a lineout and a kick at goal. Blue regions indicate field positions where the lineout has higher expected points, while red regions indicate positions where the kick at goal is preferred. Panels (a)--(f) correspond respectively to assumed gains to touch of $d_{\text{touch}} \in \{0, 5, 10, 15, 20, 25\}$\,m. The color scale reports $\Delta EP = EP_{\text{lineout}} - EP_{\text{kick}}$.}
  \label{fig:kick_lineout_all}
\end{figure}

By mapping out $\Delta EP(x, y; d_{\text{touch}})$ across the field, we obtain decision maps that show how location and assumed meters gained shape the optimal choice. Figure~\ref{fig:kick_lineout_all} displays these maps for a range of values of $d_{\text{touch}}$.

As expected, increasing $d_{\text{touch}}$ shifts the decision surface toward the lineout option, with a larger share of the pitch turning blue in Figure~\ref{fig:kick_lineout_all}.

\section{Results}

\subsection{Scenario Analysis}

Varying game context covariates in the regression model yields scenario-specific decision surfaces.

\subsubsection{Manpower Advantages and Disadvantages}

In rugby union, manpower differentials can arise following yellow or red cards. This is represented by the \texttt{Card\_Diff} covariate in the lineout EP model.

\begin{figure}[htbp!]
  \centering
  \begin{subfigure}[c]{0.32\linewidth}
    \includegraphics[width=\linewidth]{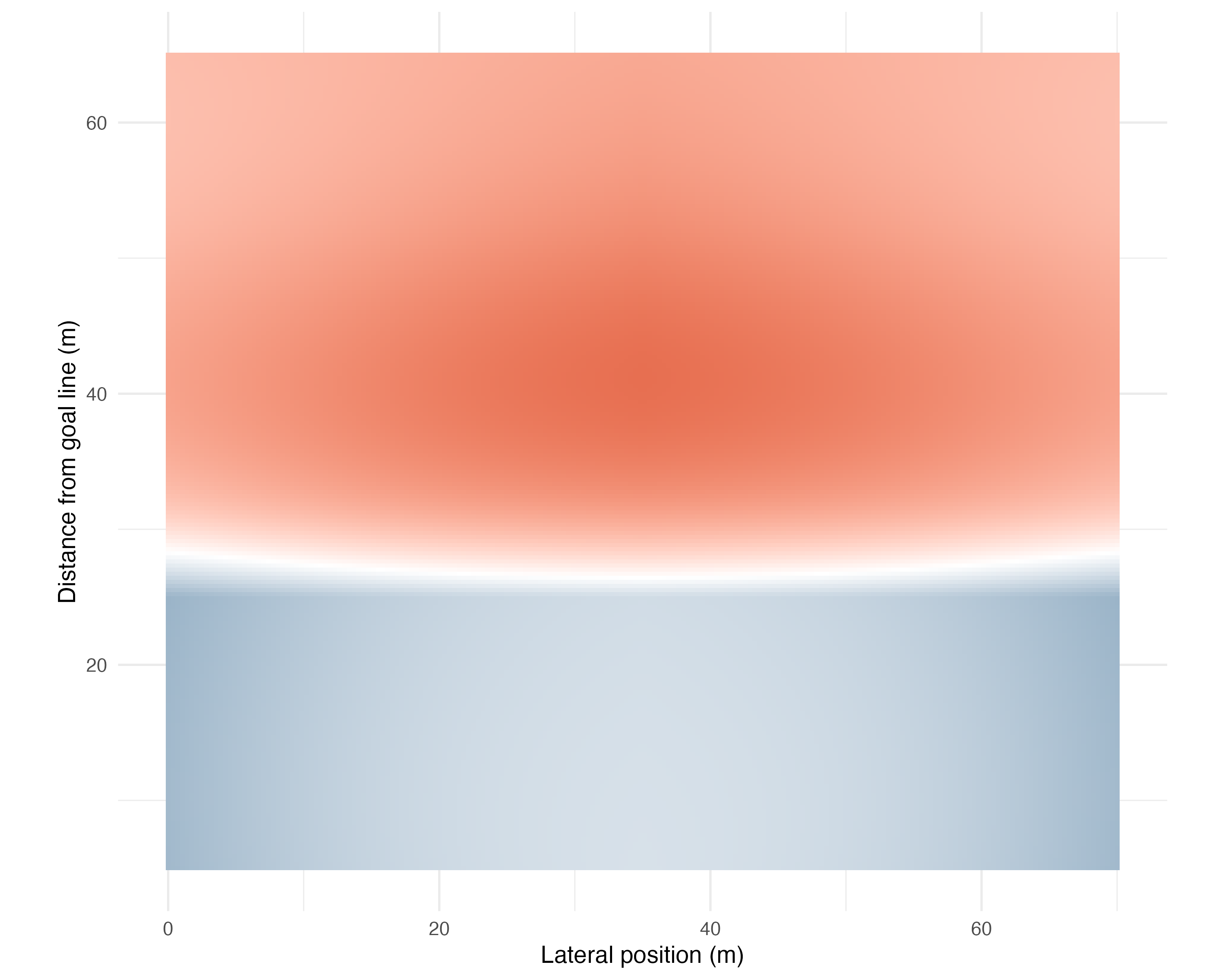}
    \caption{Attacking team has a yellow card (\texttt{Card\_Diff} $= -1$).}
  \end{subfigure}
  \hfill
  \begin{subfigure}[c]{0.32\linewidth}
    \includegraphics[width=\linewidth]{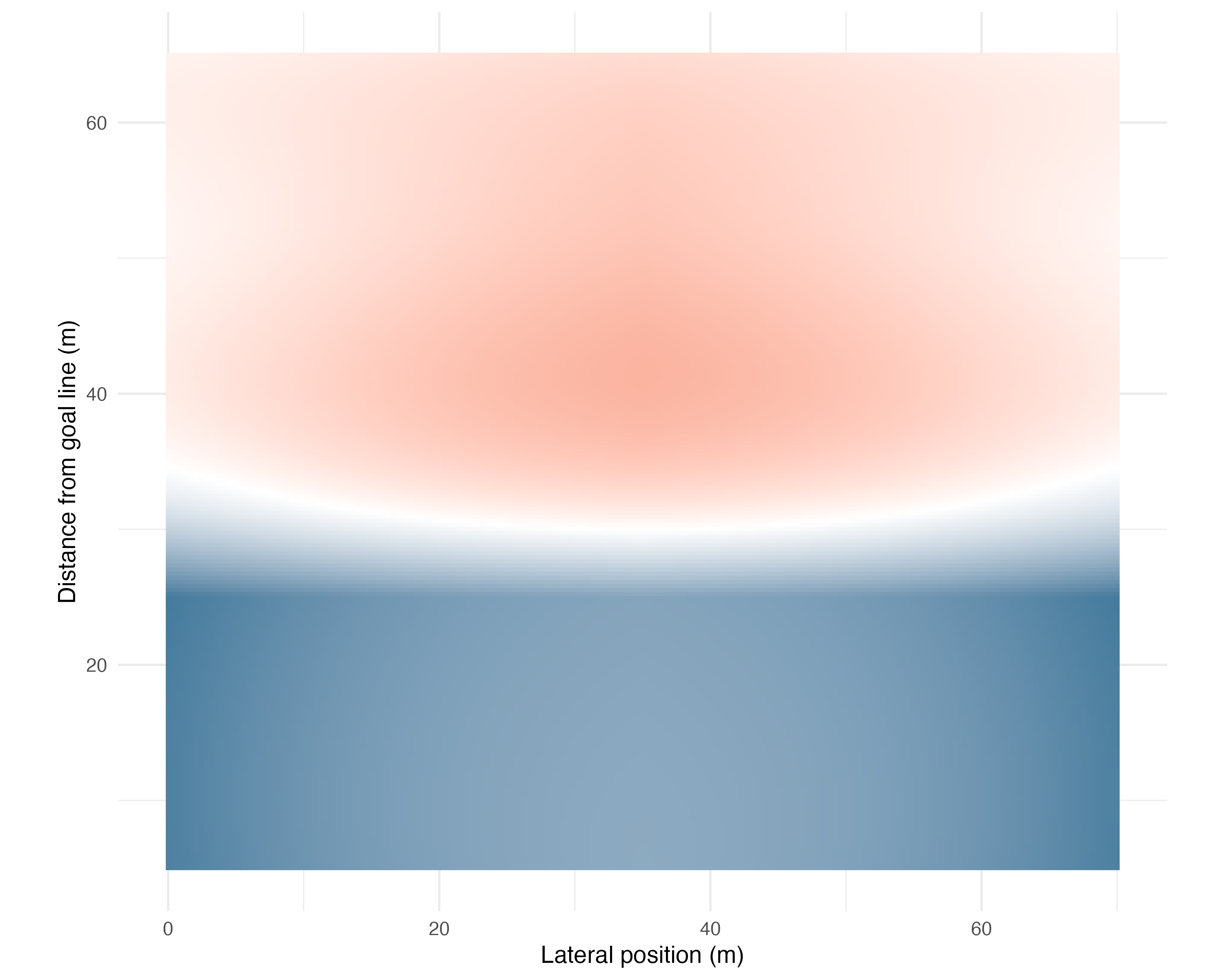}
    \caption{Neither team has a yellow card (\texttt{Card\_Diff} $= 0$).}
  \end{subfigure}
  \hfill
  \begin{subfigure}[c]{0.32\linewidth}
    \includegraphics[width=\linewidth]{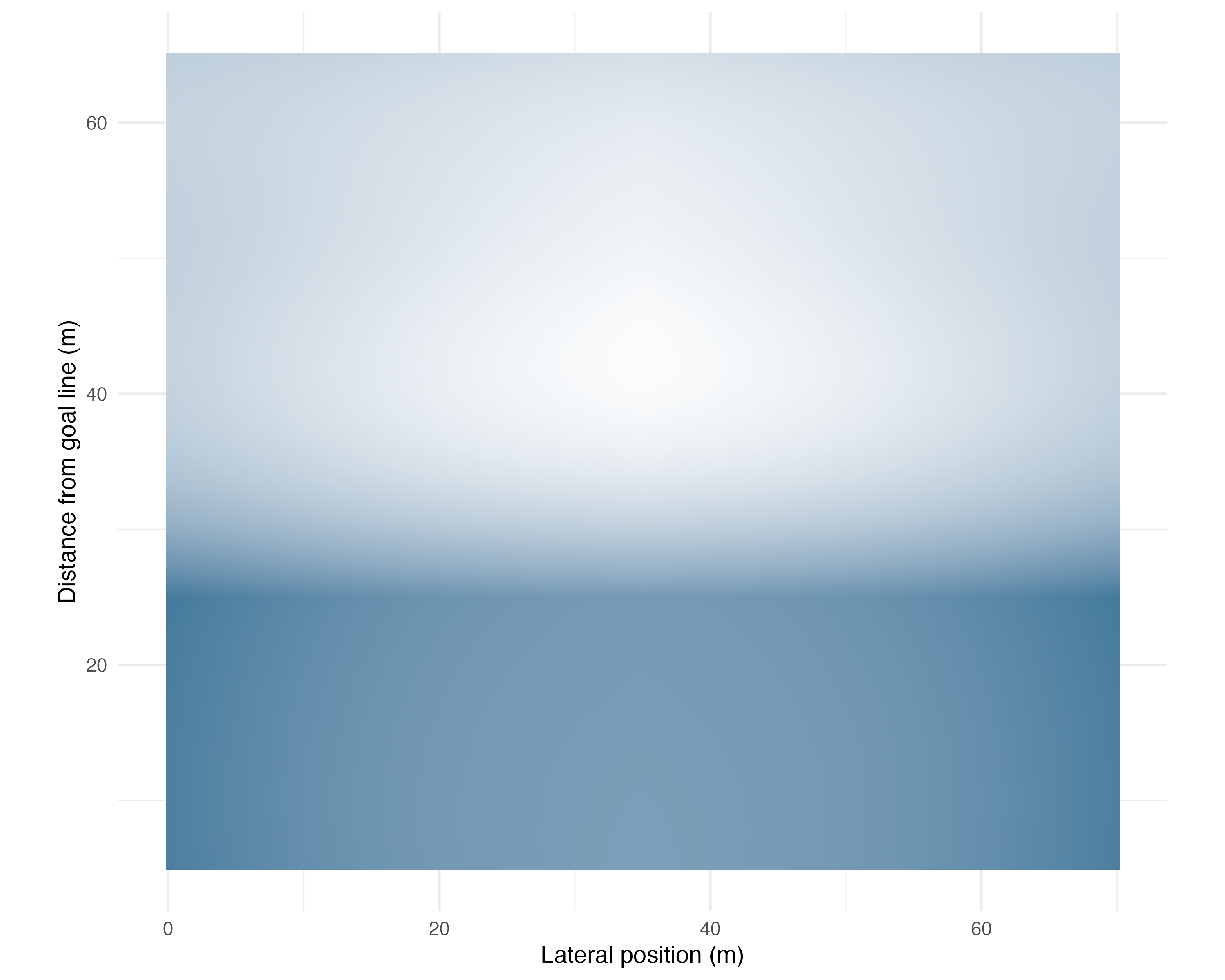}
    \caption{Defending team has a yellow card (\texttt{Card\_Diff} $= 1$).}
  \end{subfigure}
  \vspace{0.5em}
  \begin{subfigure}[c]{\linewidth}
    \centering
    \includegraphics[width=0.6\linewidth]{kick_vs_lineout_legend.png}
  \end{subfigure}
  \caption{Decision maps of $\Delta EP$ across the field under three manpower states. Panels (a)--(c) correspond to the attacking team down one player, equal manpower, and the defending team down one player.}
  \label{fig:manpower}
\end{figure}

Figure~\ref{fig:manpower} shows substantial shifts in the decision surface by yellow-card state: kicks are preferred more often when the attacking team is down a player, while lineouts are preferred more often when the attacking team has a numerical advantage.

\subsubsection{Strong versus Weak Teams}

The same approach can be used to examine how the decision surface changes with positive or negative win-percentage differentials by varying \texttt{WinPct\_Diff}.

\begin{figure}[htbp!]
  \centering
  \begin{subfigure}[c]{0.32\linewidth}
    \includegraphics[width=\linewidth]{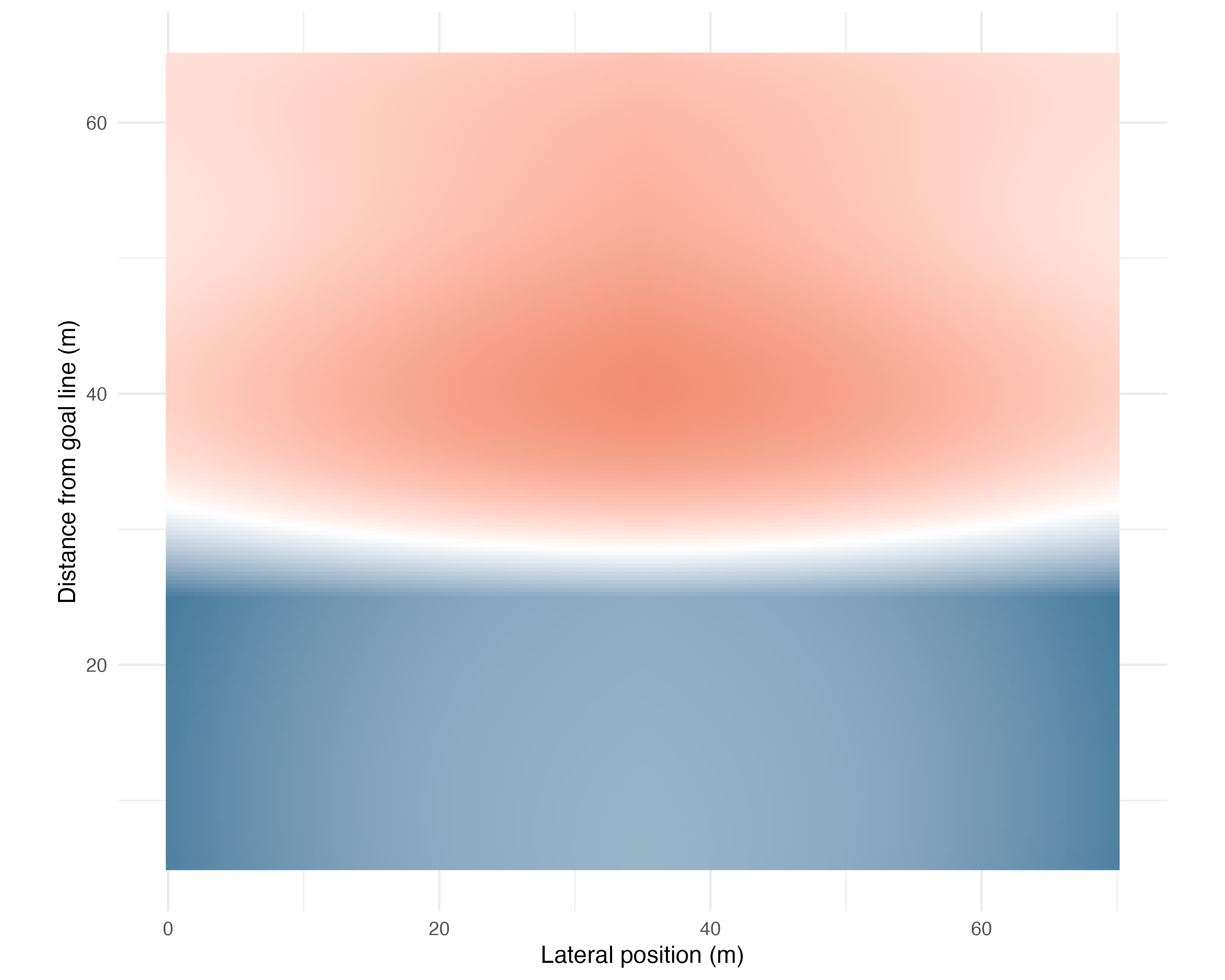}
    \caption{Attacking team is weaker (\texttt{WinPct\_Diff} $= -0.25$).}
  \end{subfigure}
  \hfill
  \begin{subfigure}[c]{0.32\linewidth}
    \includegraphics[width=\linewidth]{no_yellow.png}
    \caption{Teams are evenly matched (\texttt{WinPct\_Diff} $= 0$).}
  \end{subfigure}
  \hfill
  \begin{subfigure}[c]{0.32\linewidth}
    \includegraphics[width=\linewidth]{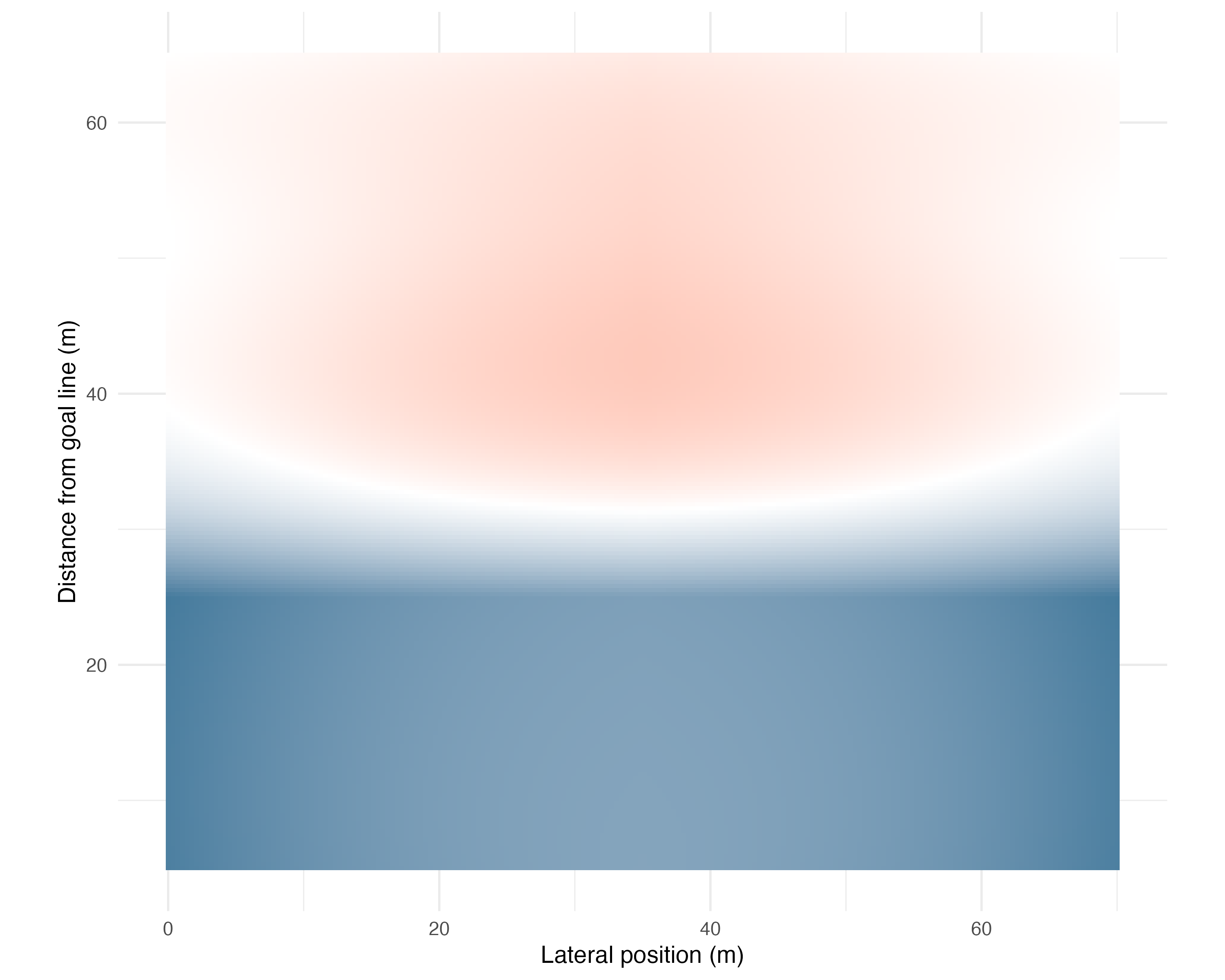}
    \caption{Attacking team is stronger (\texttt{WinPct\_Diff} $= +0.25$).}
  \end{subfigure}
  \vspace{0.5em}
  \begin{subfigure}[c]{\linewidth}
    \centering
    \includegraphics[width=0.6\linewidth]{kick_vs_lineout_legend.png}
  \end{subfigure}
  \caption{Decision maps of $\Delta EP$ across the field under three win-percentage differentials. Panels (a)--(c) correspond to the attacking team being weaker than, evenly matched with, and stronger than the defending team.}
  \label{fig:team-strength}
\end{figure}

Figure~\ref{fig:team-strength} indicates that stronger teams (positive \texttt{WinPct\_Diff}) realize larger expected points from lineouts across more of the pitch, making the aggressive option more attractive. Weaker teams (negative \texttt{WinPct\_Diff}) are relatively better off taking points from penalty kicks, though the effect is much less pronounced than the impact of yellow cards. Interestingly, the decision boundary doesn't appear to shift much when changing relative team strength.

\subsection{Case Study: New Zealand versus South Africa}

We apply the model to a real match to illustrate both single-decision analysis and aggregate decision quality. The case-study data were collected by the authors from New Zealand versus South Africa, played on September 6, 2025.

Throughout this case study, we report the \emph{mean-EP-optimal} decision (the larger of the two bootstrap-mean EP values) and two regret metrics for each decision $j$:
\[
R_{\text{mean},j}=\max\!\left\{EP^{\text{mean}}_{\text{best},j}-EP^{\text{mean}}_{\text{actual},j},\,0\right\},
\qquad
R_{\text{boot},j}=\mathbb{E}_b\!\left[\max\!\left\{EP^{(b)}_{\text{best},j}-EP^{(b)}_{\text{actual},j},\,0\right\}\right].
\]
$R_{\text{mean}}$ is deterministic from the reported mean EP values and equals the points left on the table when comparing only the two bootstrap-mean options, while $R_{\text{boot}}$ is uncertainty-aware, averaging points left on the table across bootstrap draws, and can remain positive even when the actual decision matches the mean-EP-optimal decision.

\subsubsection{A Single Penalty Decision}
\label{sec:single-penalty}

Consider a penalty awarded to South Africa in the 60th minute, 16\,m from the left touchline and 40\,m from the opposition try line. We evaluate this decision by varying the expected territory gained from a kick to touch. There were no yellow or red cards at the time. We set win-percentage differential to zero (teams treated as evenly matched), and the under-2-minute indicator is 0.

\begin{figure}[htbp!]
    \centering
    \includegraphics[width=0.75\linewidth]{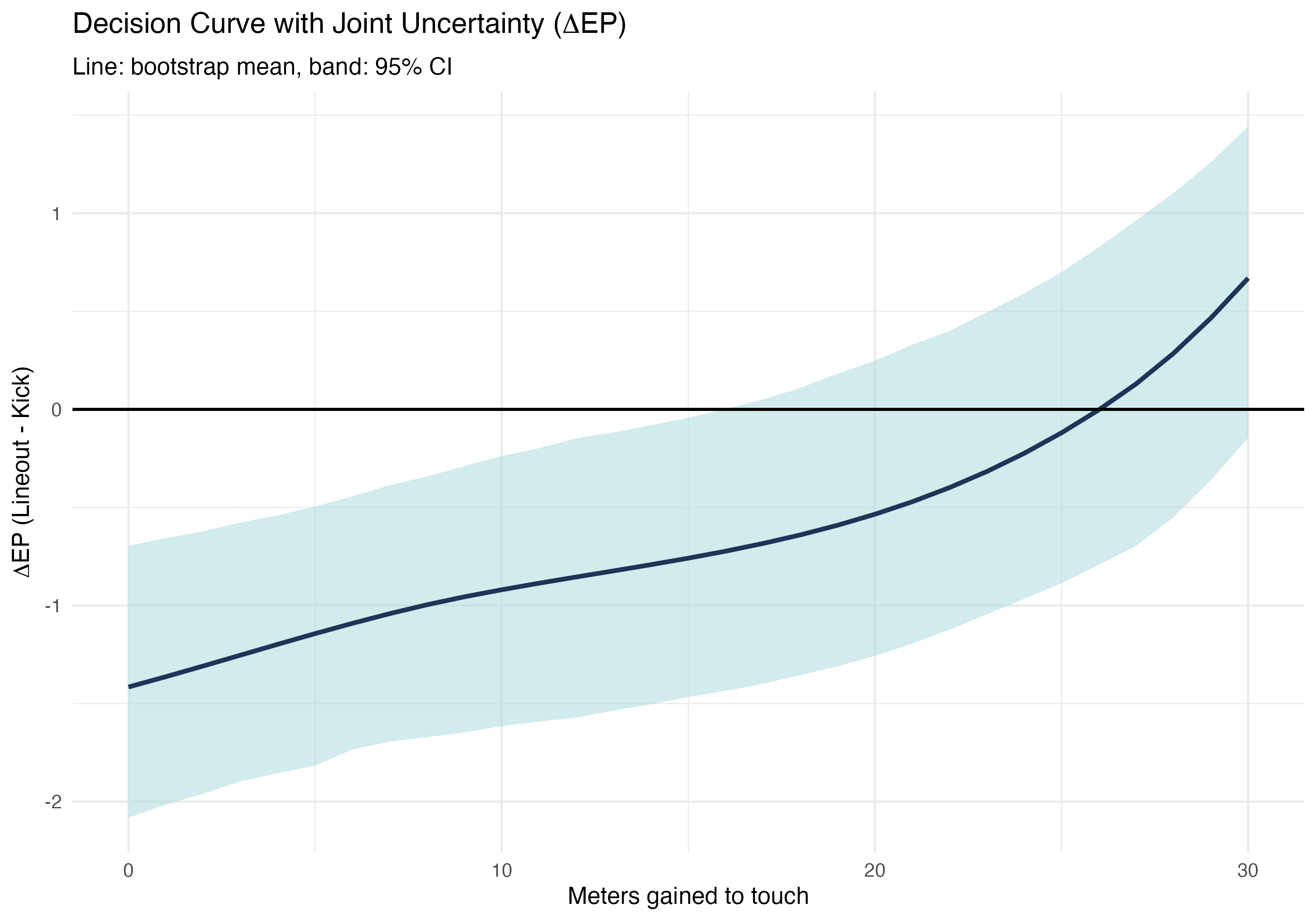}
    \caption{Expected-points difference between the lineout and kick-at-goal options as a function of meters gained to touch. The curve shows the bootstrap mean, and the shaded band shows the 95\% confidence interval.}
    \label{fig:delta-intercept-reg}
\end{figure}

Figure~\ref{fig:delta-intercept-reg} is the primary decision-uncertainty display for this marker: it plots bootstrap-mean $\Delta EP(x, y; d_{\text{touch}})$ versus $d_{\text{touch}}$ with bootstrap 95\% bands. The bootstrap-mean curve crosses zero around 26\,m.

\begin{figure}[htbp!]
  \centering
  \includegraphics[width=0.75\linewidth]{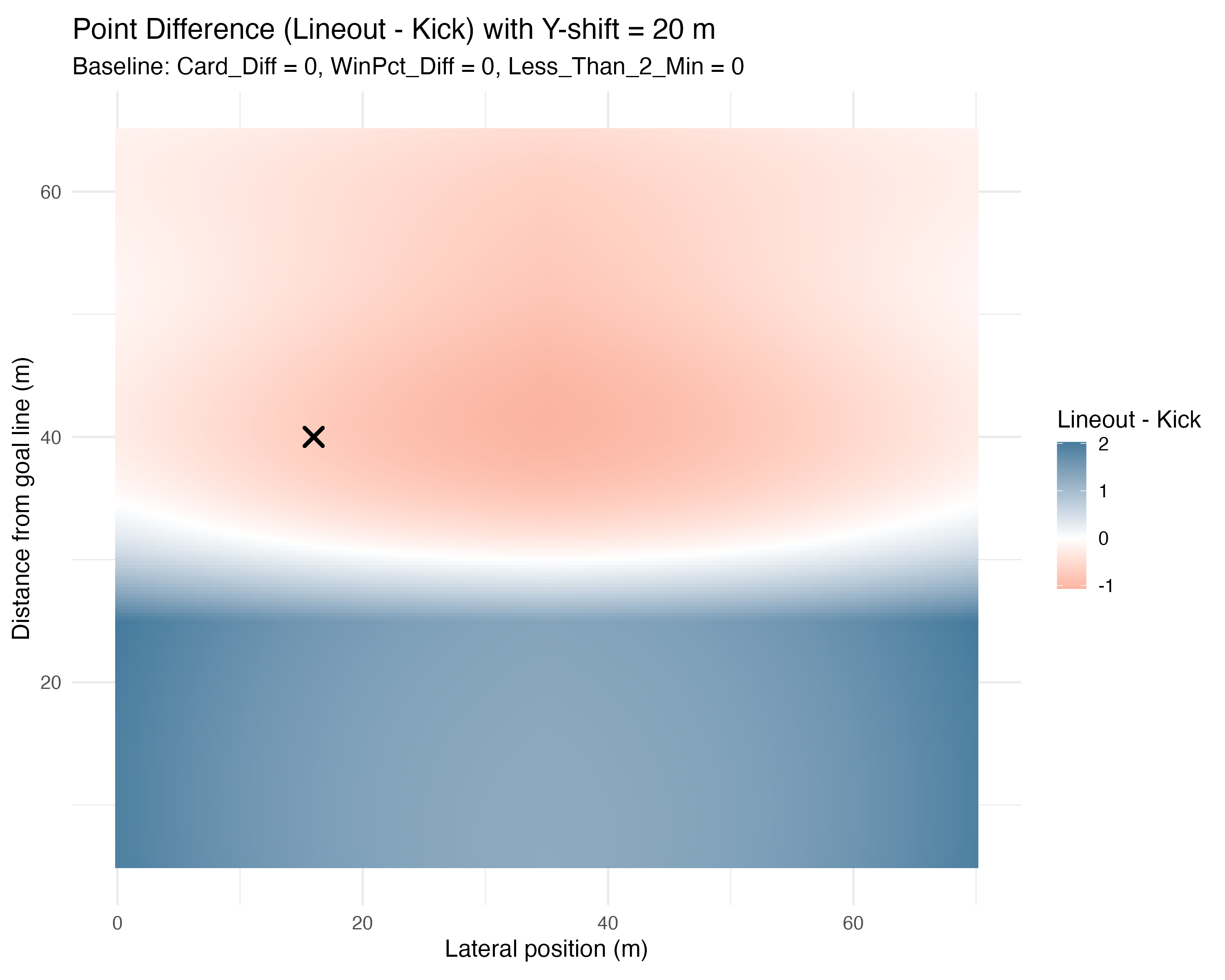}
  \caption{Penalty decision surface for the case-study decision; the marker denotes the penalty location.}
  \label{fig:my_image}
\end{figure}

In the game, South Africa gained 20\,m when they kicked to touch from this position. The model's recommendation at the penalty location, with a shift of 20\,m is shown in Figure~\ref{fig:my_image}.

At this point on the pitch, the bootstrap-mean optimal decision is to kick at goal. Table~\ref{tab:case-study} reports bootstrap means, 95\% confidence intervals, and decision-quality metrics at the marker location.

\begin{table}[htbp!]
\centering
\caption{Single-decision summary at the case-study marker ($d_{\text{touch}}=20$\,m), including bootstrap uncertainty and regret metrics for the observed lineout choice.}
\label{tab:case-study}
\begin{tabular}{lccc}
\toprule
 & \textbf{EP Lineout} & \textbf{EP Kick} & $\boldsymbol{\Delta EP}$ \\
\midrule
\textbf{Bootstrap mean} & 1.69 & 2.22 & -0.53 \\
\textbf{95\% CI} & [1.25, 2.20] & [1.65, 2.76] & [-1.26, 0.25] \\
\midrule
\multicolumn{4}{l}{\textbf{Observed decision:} lineout} \\
\multicolumn{4}{l}{$R_{\text{mean}} = 0.53,\; R_{\text{boot}} = 0.55,\; \Pr(\Delta EP > 0) = 0.08$} \\
\bottomrule
\end{tabular}
\end{table}

\begin{figure}[htbp!]
    \centering
    \includegraphics[width=0.65\linewidth]{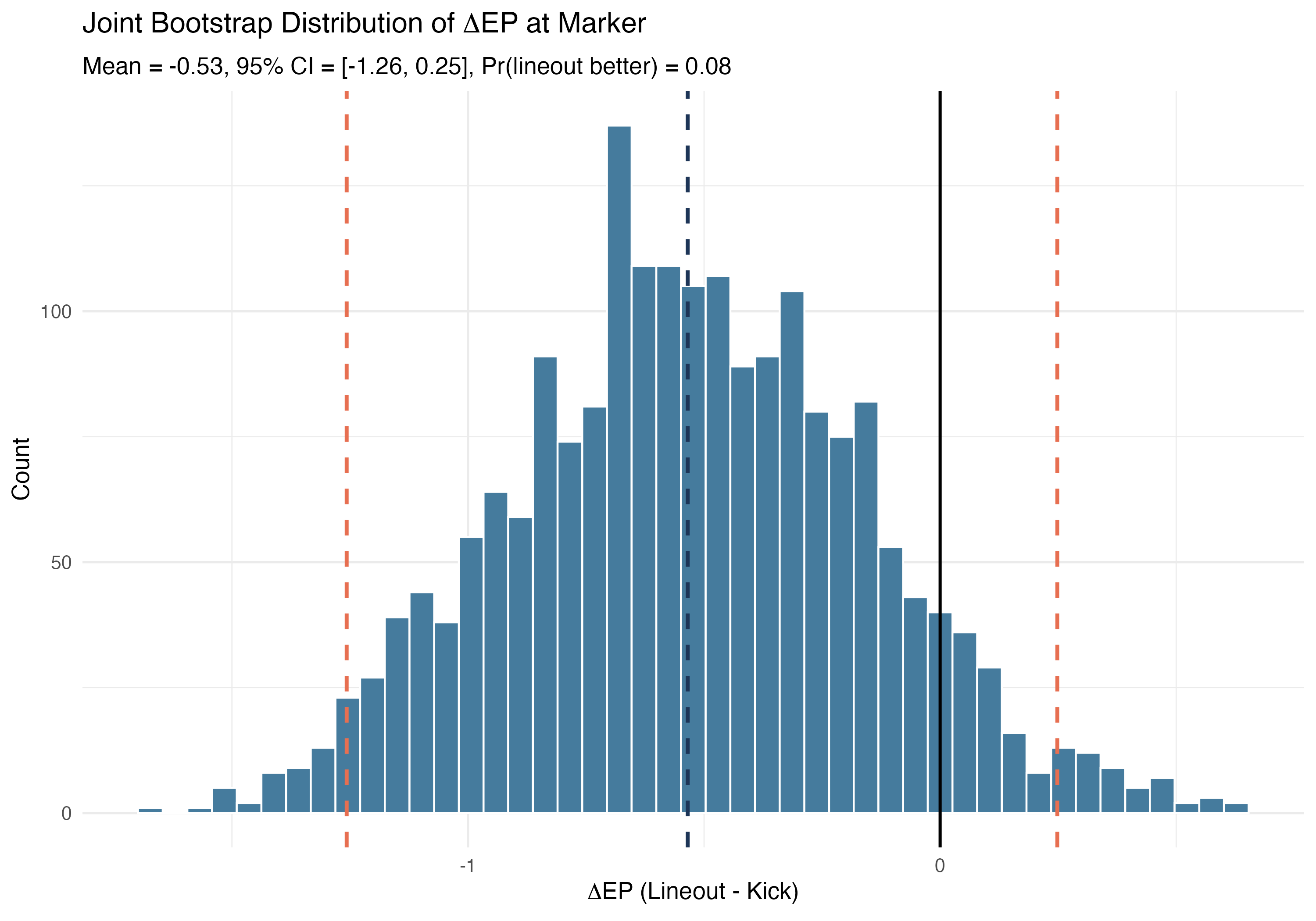}
    \caption{Joint-bootstrap distribution of $\Delta EP$ at the case-study field position for $d_{\text{touch}}=20$\,m. The black dashed line indicates the bootstrap mean, the red dashed lines show the lower and upper bounds of the 95\% confidence interval for the mean, and the solid black line marks $\Delta EP = 0$.}
    \label{fig:delta-bootstrap-marker}
\end{figure}

For this decision at $d_{\text{touch}}=20$\,m, the joint bootstrap gives mean $\Delta EP=-0.53$ with 95\% CI $[-1.26,0.25]$ and $\Pr(\Delta EP>0)=0.08$. This indicates reasonable certainty, with the bootstrap distribution favoring a kick at goal in this situation. Figure~\ref{fig:delta-bootstrap-marker} provides a complementary distributional view at this single $d_{\text{touch}}$ value.

\subsubsection{All Penalty Decisions in the Match}

Decision quality can also be compared across all relevant penalties in the match. Using the definitions above, Table~\ref{tab:ep_comparison} reports bootstrap-mean option values, the mean-EP-optimal decision, and both regret metrics for all penalties within 60\,m of the opposition try line. We assume a flat 20\,m gain for each kick to touch, treat the teams as equally matched, and use the observed under-2-minute indicator for each decision.
If the actual decision matches the mean-EP-optimal decision, then $R_{\text{mean}}=0$ by construction; $R_{\text{boot}}$ can still be positive when a non-trivial share of bootstrap draws favors the other option.

\begin{table}[H]
\centering
\small  
\caption{Bootstrap-mean EP values and two regret definitions for all case-study penalties.}
\label{tab:ep_comparison}
\resizebox{\textwidth}{!}{
\begin{tabular}{lccccccc}
\hline
Team & Actual & Mean-EP Optimal & Lineout EP & Kick EP & $\Delta EP$ & $R_{\text{mean}}$ & $R_{\text{boot}}$ \\
\hline
NZ & kick    & kick    & 1.46 & 1.96 & -0.50 & 0.00 & 0.02 \\
NZ & lineout & kick    & 1.91 & 2.19 & -0.28 & 0.28 & 0.35 \\
SA & lineout & kick    & 1.34 & 1.92 & -0.59 & 0.59 & 0.59 \\
SA & kick    & lineout & 2.89 & 2.57 &  0.32 & 0.32 & 0.36 \\
SA & kick    & kick    & 1.54 & 1.87 & -0.33 & 0.00 & 0.05 \\
SA & lineout & kick    & 0.90 & 1.66 & -0.76 & 0.76 & 0.77 \\
NZ & kick    & lineout & 3.12 & 2.71 &  0.41 & 0.41 & 0.42 \\
SA & lineout & kick    & 1.69 & 2.22 & -0.53 & 0.53 & 0.55 \\
NZ & lineout & lineout & 4.26 & 2.57 &  1.69 & 0.00 & 0.00 \\
NZ & lineout & lineout & 4.26 & 2.68 &  1.57 & 0.00 & 0.00 \\
SA & lineout & lineout & 2.69 & 2.55 &  0.14 & 0.00 & 0.10 \\
SA & lineout & kick    & 0.81 & 1.31 & -0.49 & 0.49 & 0.51 \\
SA & lineout & kick    & 1.82 & 2.89 & -1.07 & 1.07 & 1.07 \\
\hline
\end{tabular}
}
\end{table}

\begin{table}[H]
\centering
\caption{Summary metrics of penalty-decision quality for the case-study match.}
\label{tab:summary_metrics}
\begin{tabular}{lc}
\hline
Metric & Value \\
\hline
Total $R_{\text{mean}}$ & 4.46 \\
Total $R_{\text{boot}}$ & 4.80 \\
Proportion Mean-EP Optimal & 0.38 \\
\hline
\end{tabular}
\end{table}

Summary metrics of decision quality are summarized in Table~\ref{tab:summary_metrics}, indicating that the mean-EP-optimal decision was made 38\% of the time. Total regret is 4.46 points on the mean scale and 4.80 points under the uncertainty-aware bootstrap definition. For context, the game ended 24--17 to New Zealand, so this gap is non-trivial. It is important to note that South Africa was trailing by more than 3 points towards the end of the match so the final `inefficient' EP decisions were necessary to attempt to win the game.

\section{Discussion}

\subsection{Conclusions}

This paper develops a decision-ready Expected Points framework for evaluating whether to kick for goal or kick to touch following a penalty in rugby union. By combining a lineout-based EP model from phase-level Premiership data with an angle--distance model of penalty kick success and a continuation value for missed kicks, we obtain two context-aware EP surfaces and a derived decision surface that identifies where each option maximizes expected return.

The resulting decision maps make the penalty decision explicit as a function of field location, expected meters gained to touch, player advantages, and team-strength differentials. Scenario analyses show how the indifference frontier shifts with yellow cards and team quality, while the case study of New Zealand versus South Africa demonstrates how the framework can be used to evaluate both individual choices and aggregate decision quality through a regret measure. Under the smoothed continuation specification, the model still identifies inefficiencies, but many decisions sit nearer the indifference frontier than under a purely stepwise continuation mapping.

Because the framework is modular, it can be adapted to different competitions or tailored to specific teams by refitting the underlying EP components with alternative data sources. The lineout and kicking modules can be updated independently as richer data become available, providing a transparent foundation for decision-support tools in both professional and amateur settings.

\subsection{Limitations and Future Directions}

Several assumptions and data limitations should be acknowledged, along with concrete directions for future work.

First, the model combines phase-level data from club competitions with penalty-kicking data from international matches, introducing potential inconsistencies in player quality, decision-making, and context. Future work could develop fully competition-specific models by collecting phase-level and kicking data from the same league and year, or by fitting hierarchical models that allow competition-level differences in both lineout and kicking performance to be estimated explicitly.

Second, the analysis restricts team choices to kicking for goal or kicking to touch, even though teams occasionally opt for a scrum or a quick tap---particularly near the try line---albeit infrequently. Extending the framework to additional decision branches is natural: EP models can be trained for scrums and tap-and-go restarts using the same phase-level machinery, enabling multi-armed decision maps that compare all realistic options from a given penalty location.

Third, the analysis is limited to a single season, which constrains the sample size and reduces the generalizability of the findings across competitions or years. Future work could pool multiple seasons and competitions, possibly with random effects for season and league, to increase sample size, stabilize estimates in rare field zones, and quantify between-competition variation in EP surfaces.

Fourth, selection bias may influence the results \citep{BrillYurkoWyner2025}. Stronger teams are both more likely to generate attacking lineouts in the opposition half and more likely to earn penalties. As a result, lineouts in advanced field positions are disproportionately taken by stronger sides, potentially inflating expected points estimates for lineouts near the try line. A similar dynamic is present for the kicking data, since stronger sides will have generated more of the observed penalties, and coaches may only elect to kick when they deem the attempt makeable for their kicker. Future research could focus on further de-biasing EP estimates with respect to the underlying decision processes that generated the data.

Fifth, the data do not track realized meters gained when kicking to touch (or individual kicker ability for touch-finding), making it necessary to assume a translation distance between the penalty location and the resulting lineout. As a result, our decision maps are conditional on fixed $d_{\text{touch}}$ scenarios rather than integrated over an empirical $d_{\text{touch}}$ distribution. Future analyses trained on richer tracking or event data could estimate $P(d_{\text{touch}}\mid\text{context})$ directly and evaluate lineout value via iterated expectation, while also supporting player- and team-specific models for both goal-kicking accuracy and touch gain.

Sixth, the current datasets lack the granularity of precise $(x, y)$ field coordinates for lineouts, relying instead on zonal encodings and an approximate mapping to continuous coordinates. Improvements in data collection---for example, through optical or GPS tracking---would allow EP surfaces to be estimated directly on a fine spatial grid, supporting smoother decision boundaries, more accurate modeling of angle-dependent effects, and better treatment of corner and touchline situations.

Finally, this modeling framework optimizes expected points. A common critique of expected-points-based decision models is that they maximize expected points rather than expected win probability (the quantity that ultimately matters in a game). A natural extension is to develop win-probability models that incorporate field location, score differential, time remaining, and player imbalances resulting from yellow or red cards. Decisions could then be evaluated directly on the scale of win probability, with EP retained as an interpretable intermediate quantity.

\subsection{Reproducibility}

All code used to process the data, fit the models, and generate the figures and tables in this article is available at \url{https://github.com/WhartonSABI/rugby-ep}.

\subsection*{Acknowledgments}

We thank Professor Abraham J. Wyner for helpful discussion.

\subsection*{Declarations}

\textbf{Research ethics and informed consent.} Both were inapplicable.

\textbf{Author contributions.} Both authors approved the journal submission and accepted responsibility for its content.

\textbf{Use of AI tools.} The authors declared no use of large language models, AI, or machine learning tools in the journal publication.

\textbf{Conflict of interest and funding.} The authors reported no conflicts of interest and no research funding.

\textbf{Data availability.} The journal article identifies public sources for the \href{https://pubmed.ncbi.nlm.nih.gov/24598404/}{penalty-kicking data} and \href{https://journals.sagepub.com/doi/10.1177/22150218251365220}{phase-level data}.

\textbf{License.} This manuscript is distributed under the \href{https://creativecommons.org/licenses/by/4.0/}{Creative Commons Attribution 4.0 International License}. The published article is available at \href{https://doi.org/10.1515/jqas-2025-0183}{doi:10.1515/jqas-2025-0183}.

\appendix
\begin{center}
\Large\bfseries Appendix
\end{center}

\section{Model Specification Sensitivity}
\label{app:model-sensitivity}

\begin{table}[htbp]
\centering
\caption{Out-of-sample sensitivity checks for lineout and kick model specifications (grouped 5-fold CV).}
\label{tab:model-sensitivity}
\resizebox{\linewidth}{!}{%
\begin{tabular}{llccc}
\toprule
Component & Model & CV log loss & CV Brier & $\Delta$ log loss vs baseline \\
\midrule
Kick success & Logistic: angle + distance & 0.5184 & 0.1754 & 0.0000 \\
Kick success & GAM: s(angle) + s(distance) & 0.5164 & 0.1748 & -0.0020 \\
Lineout multinomial & Baseline: meter + Card + WinPct + <2min & 1.7513 & 0.7852 & 0.0000 \\
Lineout multinomial & Baseline + Card x WinPct & 1.7675 & 0.7862 & 0.0162 \\
Lineout multinomial & Baseline with separate team strengths & 1.7426 & 0.7816 & -0.0088 \\
Lineout multinomial & Baseline + continuous time & 1.7079 & 0.7707 & -0.0434 \\
\bottomrule
\end{tabular}
}
\end{table}

\section{Win-Percentage Prior Sensitivity}
\label{app:winpct-prior-sensitivity}

This section isolates sensitivity to the pre-specified win-percentage shrinkage prior while holding the baseline lineout specification fixed.

\begin{table}[htbp]
\centering
\caption{Sensitivity of baseline lineout performance to the running win-percentage shrinkage prior $m$ (grouped 5-fold CV).}
\label{tab:winpct-prior-sensitivity}
\resizebox{0.75\linewidth}{!}{%
\begin{tabular}{lccc}
\toprule
$m$ pseudo-games & CV log loss & CV Brier & $\Delta$ log loss vs $m=2$ \\
\midrule
0.5 & 1.7543 & 0.7854 & 0.0186 \\
1 & 1.7523 & 0.7819 & 0.0165 \\
2 & 1.7357 & 0.7836 & 0.0000 \\
4 & 1.7670 & 0.7835 & 0.0313 \\
8 & 1.7308 & 0.7802 & -0.0049 \\
\bottomrule
\end{tabular}
}
\end{table}

\clearpage
\bibliographystyle{apalike}
\bibliography{references}

\end{document}